\documentclass[11pt]{article}

\usepackage[a4paper,margin=1in]{geometry}

\usepackage{amsmath,amssymb,amsthm}
\newtheorem{definition}{Definition}
\theoremstyle{definition}
\newtheorem{example}{Example}

\usepackage{booktabs,array,longtable,siunitx}

\usepackage{graphicx,xcolor}
\graphicspath{{./figures/}}
\definecolor{darkred}{RGB}{139,0,0}

\usepackage[ruled]{algorithm2e}
\usepackage{enumitem}

\usepackage{authblk}
\usepackage[normalem]{ulem}
\usepackage[labelfont=bf,font=footnotesize]{caption}
\usepackage[mathlines]{lineno}
\usepackage{placeins}
\usepackage{etoolbox}

\usepackage[authoryear,round]{natbib}
\usepackage[hidelinks]{hyperref}
\usepackage[nameinlink,noabbrev]{cleveref}

\newcommand{\1}{\mathbf{1}}
\newcommand{\Var}{\operatorname{Var}}

\title{Sequential Monte Carlo Sampling of Feasible Regions Represented by R-Functions}

\author[1]{Sinan Yıldırım}
\author[2]{Sergei Kucherenko\thanks{Corresponding author. Email: \href{mailto:s.kucherenko@imperial.ac.uk}{s.kucherenko@imperial.ac.uk}}}

\affil[1]{Sabancı University, T\"urkiye}
\affil[2]{Imperial College London, UK}

\date{}

\begin{document}
\maketitle

\begin{abstract}
Feasible-region sampling and volume estimation become increasingly inefficient when constraints define only a small fraction of a convenient bounding domain. We combine R-functions, which encode compound feasibility as a continuous signed field, with sequential Monte Carlo (SMC) samplers. The proposed construction exploits not only the sign of the R-function, which determines exact feasibility, but also its continuous negative values outside the feasible set, which provide graded information on constraint violation. Logistic intermediate distributions use this information to guide particles progressively towards feasibility, while the final target distribution remains uniform on the feasible region.

Experiments on intersections and unions, including disconnected cases of hyperballs, show substantial variance reductions relative to direct rejection sampling, with the statistical advantage generally increasing as the feasible fraction decreases. For sufficiently small feasible regions, these advantages persist even after accounting for computational cost. To isolate the benefit of continuous R-function information, we also compare the proposed method with an SMC sampler with a region-only (RO) construction that uses the same sequential particle machinery but retains only binary feasible/infeasible information. We demonstrate that the R-function-informed SMC method concentrates particles in the feasible region earlier and produces lower-variance volume estimates than RO.

\end{abstract}

\section{Introduction}
\label{sec:introduction}

Sampling from constrained regions and estimating their volumes are fundamental tasks in uncertainty analysis, reliability analysis, and process systems engineering \citep{Au_and_Beck_2001, kucherenko2025rfunctions, boukouvala2012}. A simple approach is direct rejection sampling: points are generated uniformly from a convenient bounding domain and retained when they satisfy all constraints. Although straightforward and exact, rejection sampling uses only binary membership information, and its efficiency deteriorates in direct proportion to the feasible fraction of the bounding domain. When this fraction becomes small, very large numbers of samples may therefore be required to obtain an adequate representation of the feasible region or an accurate estimate of its volume.

R-functions \citep{rvachev1978theory,rvachev1982theory,shapiro1992rfunctions} provide an alternative representation of constrained regions by combining multiple inequalities into a single continuous signed function whose sign determines feasibility and whose zero level set represents the boundary \citep{kucherenko2025rfunctions}. Importantly, the R-function contains more information than a binary feasible/infeasible indicator: its continuous negative values outside the feasible region provide graded information about constraint violation. In this work, we exploit this additional information within a \emph{sequential Monte Carlo} (SMC) sampling framework \citep{delmoral2006smc}, which has been arguably underused in statistics despite its advantages \citep{Dai_et_al_2022}. In this framework, a sequence of logistic intermediate targets uses the continuous R-function to guide particles progressively towards feasibility, while the final hard target remains uniform on the feasible region. The associated SMC estimate of the normalising-constant ratio also provides an estimate of the feasible volume.

The main contribution of this work is an R-function-informed SMC construction for feasible-region sampling and volume estimation, in which continuous signed constraint information is used to define intermediate target distributions. The design choice studied here is a bridge that preserves equal target density throughout the feasible region while using signed constraint values outside it.  We use this bridge in a standard SMC sampler and compare the resulting method with rejection sampling and an SMC sampler that uses region-only intermediate distributions, including an oracle schedule based on the known volume in benchmark problems. Specifically, we assess the method on intersections and disconnected unions of hyperballs over increasing dimension. Particular attention is given to small feasible fractions and disconnected regions, where direct rejection becomes inefficient and sequential particle methods face their own challenges of computational overhead and component preservation. The experiments assess accuracy, computational cost, and particle evolution for these configurations. 

\section{R-Functions: Mathematical Background} \label{sec:R_functions}

R-functions, introduced by V.~L.~Rvachev for geometric modelling and implicit function representation \citep{rvachev1978theory,rvachev1982theory,shapiro1992rfunctions}, provide an algebraic framework for combining implicit constraints while preserving their logical structure. Their application to the identification of complex feasible regions in multidimensional spaces is described in detail by \citet{kucherenko2025rfunctions}. 

Let $\mathcal{X} \subseteq \mathbb{R}^{d}$ and let $\phi_i: \mathcal{X} \to \mathbb{R}$ be continuous functions that implicitly define constraints using the positive-inside convention,
\[
\phi_i(\mathbf{x}) \geq 0
\quad\Longleftrightarrow\quad
\mathbf{x}\ \text{satisfies constraint }i.
\]
R-functions combine these primitives so that logical operations on the constraints are reproduced by the sign of a single scalar function. R-functions are defined through operators called R-conjunction and R-disjunction.
\begin{definition}
An operator $\wedge$ is called an R-conjunction if, given any pair of constraint functions $\phi_{1}, \phi_{2}$, it satisfies 
\begin{equation} \label{eq: R-conjunction}
[\phi_{1} \wedge \phi_{2}](\mathbf{x}) \geq 0 \Leftrightarrow  \phi_{1}(\mathbf{x}) \geq 0 \text{ and } \phi_{2}(\mathbf{x}) \geq 0
\end{equation}
An operator $\vee$ is called an R-disjunction if, given any pair of constraint functions $\phi_{1}, \phi_{2}$, it satisfies
\begin{equation} \label{eq: R-disjunction}
[\phi_{1} \vee \phi_{2}](\mathbf{x}) \geq 0 \Leftrightarrow  \phi_{1}(\mathbf{x}) \geq 0 \text{ or } \phi_{2}(\mathbf{x}) \geq 0
\end{equation}
\end{definition}
Commonly used symmetric families of R-conjunction and R-disjunction operators are $\wedge_\alpha$ and $\vee_\alpha$, defined as 
\begin{align*}
\phi_1 \wedge_\alpha \phi_2
&=
\frac{1}{1+\alpha}
\left(
\phi_1+\phi_2
-
\sqrt{\phi_1^2+\phi_2^2-2\alpha\phi_1\phi_2}
\right),
\label{eq:R_conjunction_general}
\\
\phi_1 \vee_\alpha \phi_2
&=
\frac{1}{1+\alpha}
\left(
\phi_1+\phi_2
+
\sqrt{\phi_1^2+\phi_2^2-2\alpha\phi_1\phi_2}
\right),
\end{align*}
where $-1<\alpha\leq1$. In the present work we use $\alpha=1$, for which
these expressions reduce to
\[
\phi_1\wedge_1\phi_2=\min(\phi_1,\phi_2),
\qquad
\phi_1\vee_1\phi_2=\max(\phi_1,\phi_2).
\]

An R-function on $\mathcal{X}$ that serves as a signed representation of the feasible region can be obtained by combining constraint primitives with R-conjunction and R-disjunction, such as $R = \phi_{1} \vee \phi_{2}$ or $R = (\phi_{1} \wedge \phi_{2}) \vee \phi_{3}$; see \cite{Shapiro_2007} for the broader theory. Here $R: \mathcal{X} \to \mathbb{R}$ denotes the function, whereas $\mathcal{R}=\{\mathbf{x}\in\mathcal{X}:R(\mathbf{x})\geq0\}$ denotes the feasible region it represents. For example, suppose that a feasible region $\mathcal{R}$ is defined in terms of constraint functions $\phi_{1}, \phi_{2}, \phi_{3}$ as follows:
\[
\mathcal{R} = \{  \mathbf{x} \in \mathcal{X}: \phi_{1}(\mathbf{x}) \geq 0 \text{ or } (\phi_{2}(\mathbf{x}) \geq 0 \text{ and } \phi_{3}(\mathbf{x}) \geq 0) \}
\]
Then, an R-function for $\mathcal{R}$ can be constructed with a pair of R-conjunction and  R-disjunction operators $\wedge, \vee$ as 
\[
R = \phi_{1} \vee (\phi_{2} \wedge \phi_{3}).
\]
Importantly for the present work, $R(\mathbf{x})$ is not used merely as a binary feasibility test. If the constraints $\phi_{i}$ and operators $\wedge, \vee$ involved in the definition of $R$ are continuous, so is $R$. With suitably chosen constraint primitives, the continuous negative values of $R$ outside $\mathcal R$ provide graded information about constraint violation. These values do not generally represent geometric distance to the boundary; their scale depends on the primitives and R-function construction. It is this additional information, discarded by a binary feasibility indicator, that is exploited by the SMC construction developed in Section~\ref{sec:smc}.

\begin{example}[\textbf{Illustrative two-dimensional disjoint region}] \label{ex: Rfunctionexample}
We first consider Example~3 of \citet{kucherenko2025rfunctions}, originally introduced by \citet{boukouvala2012}, which provides a simple two-lobed geometry with disconnected interiors. On $x_1 \in [-2,2]$ and $x_2\in[-1,1]$, the feasible region is defined by the following constraint functions:
\begin{align*}
\phi_1(\mathbf{x})& = -4x_1^2+2.1x_1^4-\frac{x_1^6}{3}
 -x_1x_2+4x_2^2-4x_2^4\\
\phi_2(\mathbf{x})&=-2x_1+x_2+3,\\
\phi_3(\mathbf{x})&= 0.8x_1-x_2+1.8.
\end{align*}
The first constraint is based on the six-hump camel test function and gives the example its multimodal geometry. The three constraints are combined by the R-conjunction operator $\wedge_{1}$ in \eqref{eq: R-conjunction} to construct 
\[
R = \phi_{1} \wedge_{1} \phi_{2} \wedge_{1} \phi_{3}.
\]
Explicitly, the expanded R-function is
\begin{equation*}
\begin{aligned}
R(\mathbf{x})
={}&x_2^2-\frac{x_2}{4}
-\frac12\left|\frac{9x_1}{5}-\frac{3x_2}{2}
+\frac{B}{2}+\frac{x_1x_2}{2}+2x_1^2-2x_2^2
-\frac{21x_1^4}{20}+2x_2^4+\frac{x_1^6}{6}+\frac{3}{10}\right|\\
&-\frac{B}{4}-\frac{x_1x_2}{4}-x_1^2-\frac{x_1}{10}
+\frac{21x_1^4}{40}-x_2^4-\frac{x_1^6}{12}
+\frac{33}{20},
\end{aligned}
\end{equation*}
where
\begin{equation*}
B=\left|x_2-2x_1+x_1x_2+4x_1^2-4x_2^2
-\frac{21x_1^4}{10}+4x_2^4+\frac{x_1^6}{3}+3\right|.
\end{equation*}
The feasible set is exactly the nonnegative level set $R(\mathbf{x})\geq0$ and consists of two lobes with disconnected interiors, meeting at the origin. Figure~\ref{fig:rheatmap} shows how the R-function contains more information than binary membership. The left panel displays the R-function over the entire bounding region. The right panel restricts the display to $R(\mathbf{x})>0$.

\begin{figure}[ht]
\centering
\includegraphics[width=.50\linewidth]{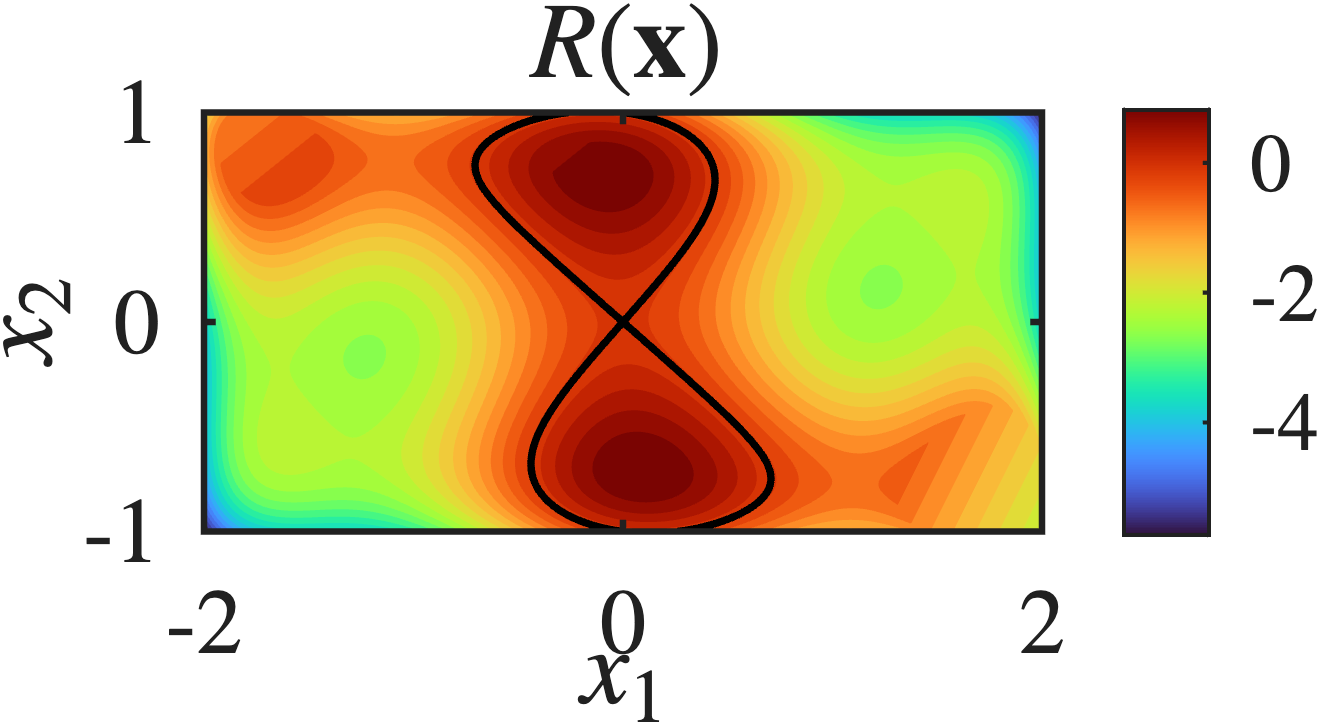}\hfill
\includegraphics[width=.50\linewidth]{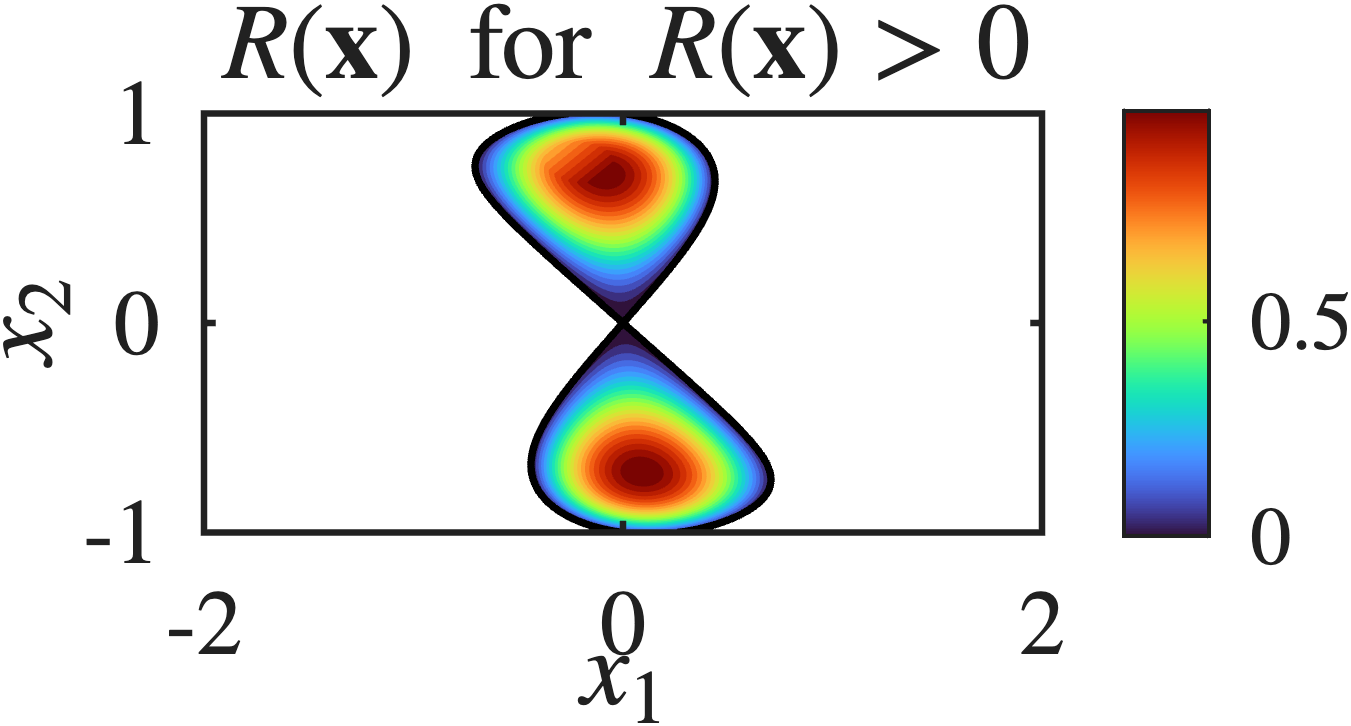}\hfill
\caption{Illustrative two-dimensional example. Left: $R$-function values over
the bounding region $\mathcal A$. Right: $R$-function values restricted to
$R(\mathbf{x})>0$, with the infeasible region shown in white.}
\label{fig:rheatmap}
\end{figure}
\end{example}

\section{Rejection sampling and volume estimation}
Let $\mathcal{A} \supseteq\mathcal R$ be a convenient bounding region with known volume
$|\mathcal{A}|=A$, and let $V = |\mathcal R|$ denote the unknown feasible volume. We will assume throughout that $A \geq V > 0$. If $X_1,\ldots,X_N$ are sampled independently and uniformly from $\mathcal A$, the
standard rejection-sampling (or indicator importance-sampling) estimator is
\begin{equation*}
\widehat V_{\rm IS} = \frac{A}{N}\sum_{i=1}^{N} \1\{X_i\in\mathcal R\}.
\end{equation*}
Since the feasibility probability is $V/A$, this estimator is unbiased in that $\mathbb{E}[\widehat{V}_{\rm IS}] =\frac{A}{N} \left( N \frac{V}{A} \right) = V$,  and has variance
\begin{equation}
\Var(\widehat V_{\rm IS}) = \frac{1}{N}V(A-V).
\label{eq:is-variance}
\end{equation}
More revealing is the relative variance,
\begin{equation*}
\Var\left(\frac{\widehat V_{\rm IS}}{V}\right) = \frac{1}{N} \left(\frac{A}{V}-1\right).
\end{equation*}
Thus, the efficiency of direct rejection sampling is governed by the inverse feasible fraction $A/V$, which can increase rapidly with the dimension $d$. When $A/V$ becomes large, increasingly many samples are required to obtain a given relative accuracy; in sufficiently rare-event regimes, a finite experiment may contain no feasible points at all.

\section{Sequential Monte Carlo Samplers} \label{sec:smc}

An SMC sampler introduces a sequence of probability distributions on $\mathcal{X}$, 
\begin{equation*}
\pi_k(\mathbf{x})=\frac{\gamma_k(\mathbf{x})}{Z_k}, \qquad k=0,\ldots,K,
\end{equation*}
where $\gamma_k$ is an unnormalised density, which is known pointwise, and
\begin{equation*}
Z_k=\int \gamma_k(\mathbf{x})\,d\mathbf{x}
\end{equation*}
is its normalising constant, which is typically unknown except for $Z_{0}$. The initial distribution $\pi_0$ is chosen so that it can be sampled easily and $Z_0$ is known, whereas the final distribution $\pi_K$ is the desired uniform distribution on the feasible region $\mathcal R$. Consecutive targets must have compatible support, specifically $\pi_{k} \ll \pi_{k-1}$, which means that $\pi_{k}$ is absolutely continuous with respect to $\pi_{k-1}$, that is, $\gamma_k(\mathbf{x})=0$ almost everywhere wherever $\gamma_{k-1}(\mathbf{x})=0$. The practical efficiency of the SMC sampler depends on constructing intermediate distributions that provide a sufficiently gradual bridge from $\pi_0$ to $\pi_K$. The SMC sampler provides
\begin{itemize}
\item A particle approximation $\{ X_{K}^{(i)}\}_{i = 1}^{N}$ of the final target distribution $\pi_{K}$.
\item An unbiased estimator of $Z_{K}/Z_{0}$ \citep{delmoral2006smc,Dai_et_al_2022}.
\end{itemize} 

The unbiasedness statement assumes a fixed target sequence with finite, positive normalising constants and compatible supports, initial particles drawn independently from $\pi_0$, conditionally unbiased resampling, and mutation kernels that leave each $\pi_k$ invariant. If all incremental weights vanish, the normalising-constant estimate is zero and the run terminates without a valid final particle population.

The SMC sampler for a generic sequence of $\{ \gamma_{k} \}_{k = 0}^{K}$ is summarised in Algorithm~\ref{alg:smc}. At each stage, particles are reweighted to account for the change from $\pi_{k-1}$ to $\pi_k$, resampled according to their normalised weights, and subsequently moved using a Markov chain Monte Carlo (MCMC) kernel that leaves $\pi_k$ invariant. Resampling tends to replicate particles with large weights and discard particles with small weights, thereby reducing particle diversity. The subsequent MCMC move rejuvenates the resampled particles and can restore some diversity. A commonly used MCMC kernel is the random-walk Metropolis--Hastings (MH) algorithm with Gaussian proposals. Given a current particle $X$, a proposal is generated as
\begin{equation*}
X'=X+\sigma_q Z, \qquad Z\sim\mathcal N(0,I_d),
\end{equation*}
and accepted with probability
\begin{equation*}
\alpha(X,X') = \min\left\{ 1, \frac{\pi_k(X')}{\pi_k(X)} \right\} = \min\left\{ 1, \frac{\gamma_k(X')}{\gamma_k(X)} \right\};
\end{equation*}
otherwise, the proposal is rejected. Observe that the rejuvenation step can be implemented directly using the unnormalised intermediate target $\gamma_k$, because the usually unknown normalising constant $Z_k$ in $\pi_k(\mathbf{x})=\gamma_k(\mathbf{x})/Z_k$ cancels from the MH ratio. Moreover, Algorithm \ref{alg:smc} requires only the unnormalised distributions $\{ \gamma_{k} \}_{k = 0}^{K}$.

\begin{algorithm}[H]
\caption{SMC sampler for feasible-region sampling and volume estimation}
\label{alg:smc}

\textbf{Input:} Unnormalised targets $\{\gamma_k\}_{k=0}^{K}$, initial particles $\{X_0^{(i)}\}_{i=1}^{N}\sim\pi_0$, and Gaussian random-walk Metropolis--Hastings kernels targeting $\pi_k$.\\

Set $\widehat L_0=1$.

\For{ $k=1,\ldots,K$}{
\textbf{Weighting:} Compute the unnormalised weights
\begin{equation*}
w_k^{(i)} = \frac{\gamma_k(X_{k-1}^{(i)})}{\gamma_{k-1}(X_{k-1}^{(i)})}, \qquad i=1,\ldots,N.
\end{equation*}

Update the normalising-constant ratio estimate,
\begin{equation*}
\widehat L_k = \widehat L_{k-1} \left( \frac{1}{N}\sum_{i=1}^{N}w_k^{(i)} \right).
\end{equation*}

\textbf{Resampling:} Sample indices $ I_k^{(1:N)}\sim\operatorname{Resample}(w_k^{(1:N)})$ using an unbiased resampling algorithm given unnormalised weights $w_{k}^{(1:N)}$, and obtain the resampled particles as $\widetilde X_k^{(i)}=X_{k-1}^{(I_k^{(i)})}$ for $i = 1, \ldots, N$.
 
\textbf{Rejuvenation:} Move each resampled particle via an MCMC move with transition kernel $M_{k}$ targeting $\pi_k$, obtaining $X_{k}^{(i)} \sim M_{k}(\widetilde{X}_{k}^{(i)}, \cdot)$ for $i = 1, \ldots, N$.
}

\textbf{Output:} A particle approximation $\{ X_{K}^{(i)} \}_{i = 1}^{N}$ of $\pi_{K}$ and an unbiased estimator $\widehat{L}_{K}$ of the normalising-constant ratio $Z_K/Z_0$.
\end{algorithm}

\subsection{R-function-informed SMC samplers} \label{sec: R-function-informed SMC Samplers}
When SMC is to be applied to our problem of exploring a feasible region, $\pi_{K}$ can be chosen to be the uniform distribution over $\mathcal{R}$, and the initial distribution $\pi_{0}$ can be chosen as the uniform distribution over $\mathcal{A} \supseteq \mathcal{R}$ with volume $|\mathcal{A}| = A > V$ as assumed before. Thus, SMC can be used to provide a particle approximation of the uniform distribution over the feasible region and an unbiased estimator of its volume. 

The central role of the R-function in the proposed construction is that it provides more information than the binary event $\mathbf{x}\in\mathcal R$, hence it can be used to construct useful intermediate distributions. Since
\begin{equation*}
R(\mathbf{x})\geq0 \quad\Longleftrightarrow\quad \mathbf{x}\in\mathcal R,
\end{equation*}
its sign alone determines feasibility. With suitably chosen constraint primitives, the continuous negative values of $R(\mathbf{x})$ provide graded information about constraint violation outside $\mathcal R$. Although these values do not generally represent geometric distance to the boundary, they distinguish infeasible points and allow informative intermediate distributions to be constructed. This information is discarded when feasibility is represented only by a binary indicator. To exploit this information while retaining a uniform final target on
$\mathcal R$, define the function $\psi: \mathcal{X} \to \{1\} \cup (-\infty, 0)$ as
\begin{equation*}
\psi(\mathbf{x})= \begin{cases} 1, & R(\mathbf{x})\geq0,\\ R(\mathbf{x}), & R(\mathbf{x})<0. \end{cases}
\end{equation*}
The unnormalised initial, intermediate, and final targets are
\begin{equation}
\gamma_k(\mathbf{x}) = \left\{1+\exp[-\beta_k\psi(\mathbf{x})]\right\}^{-1} \1_\mathcal{A}(\mathbf{x}), \qquad k=0,\ldots,K,
\label{eq:smc-logistic-targets}
\end{equation}
where the inverse temperatures $\beta_k$ satisfy $0=\beta_{0}<\beta_1<\cdots<\beta_{K-1} < \beta_{K} = \infty$. At each finite temperature, the target is constant on the feasible region and assigns graded positive density to infeasible points according to their R-function values. 

SMC in Algorithm \ref{alg:smc} with these intermediate distributions provides both a particle approximation to the uniform distribution on $\mathcal R$ and an estimate of its volume $V = |\mathcal{R}| $. At $\beta_0=0$,
\begin{equation*}
\gamma_0(\mathbf{x})=\frac{1}{2}\1_{\mathcal{A}}(\mathbf{x}), \qquad Z_0=\frac{A}{2}.
\end{equation*}
The choice $\beta_{K} = \infty$ for the last stage yields the target distribution 
\[
\gamma_K(\mathbf{x})=\1_{\mathcal R}(\mathbf{x}), \quad Z_K = \int \1_{\mathcal R}(\mathbf{x})\,d\mathbf{x} = V.
\]
Since $\widehat L_K$ is the SMC estimator of the normalising-constant ratio $Z_K/Z_0 = 2V/A$, the SMC volume estimator is therefore
\begin{equation*}
\widehat V_{\rm SMC} = \frac{A}{2}\widehat L_K.
\end{equation*}
At the final stage, the particles $\{X_{K}^{(i)}\}_{i = 1}^{N}$ represent a particle approximation of the uniform distribution over the feasible region $\mathcal R$. 

The R-function information enters the SMC procedure at weighting and rejuvenation steps through $\{ \gamma_{k}\}_{k =0}^{K}$. Consequently, infeasible particles are not treated as equivalent: proposals towards regions of weaker constraint violation are favoured through the current intermediate target.

\begin{figure}[htb]
\centerline{
\includegraphics[width=.19\linewidth]{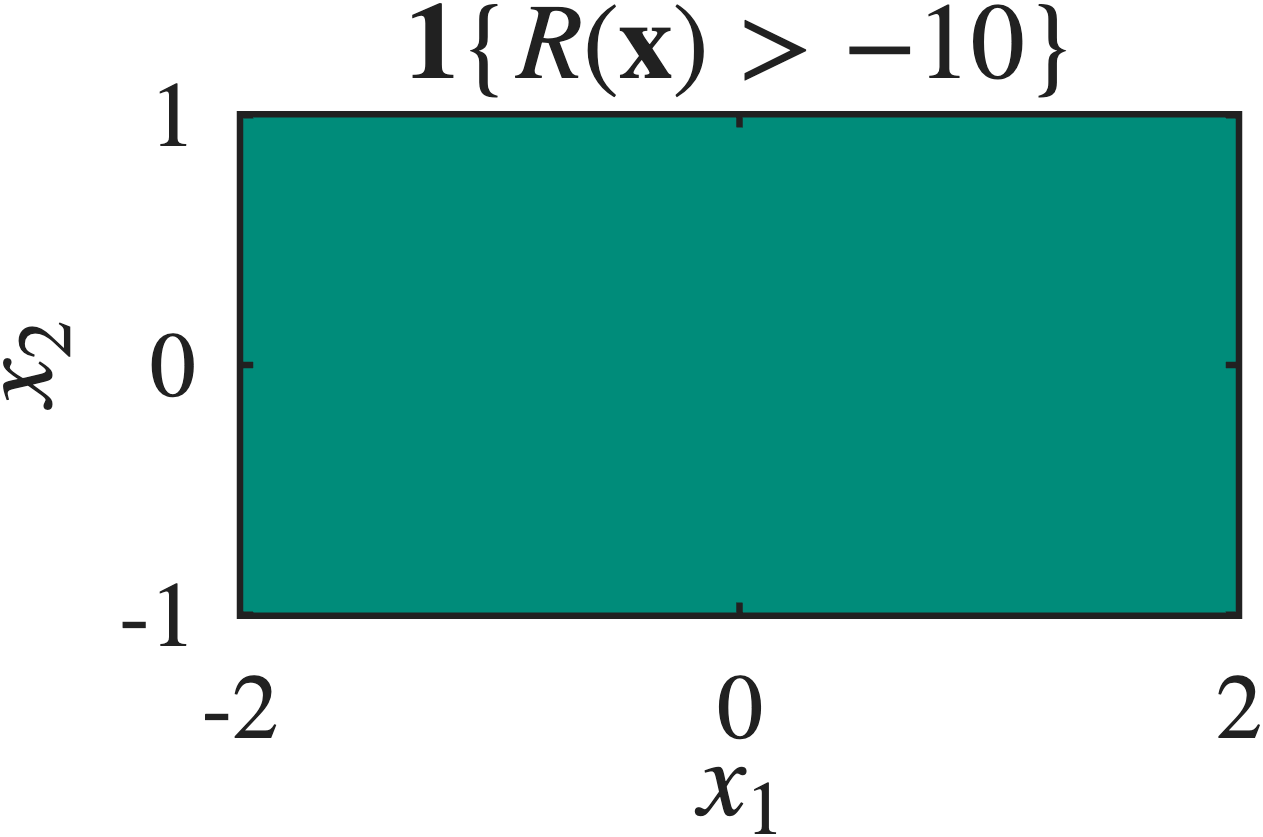}\hfill
\includegraphics[width=.19\linewidth]{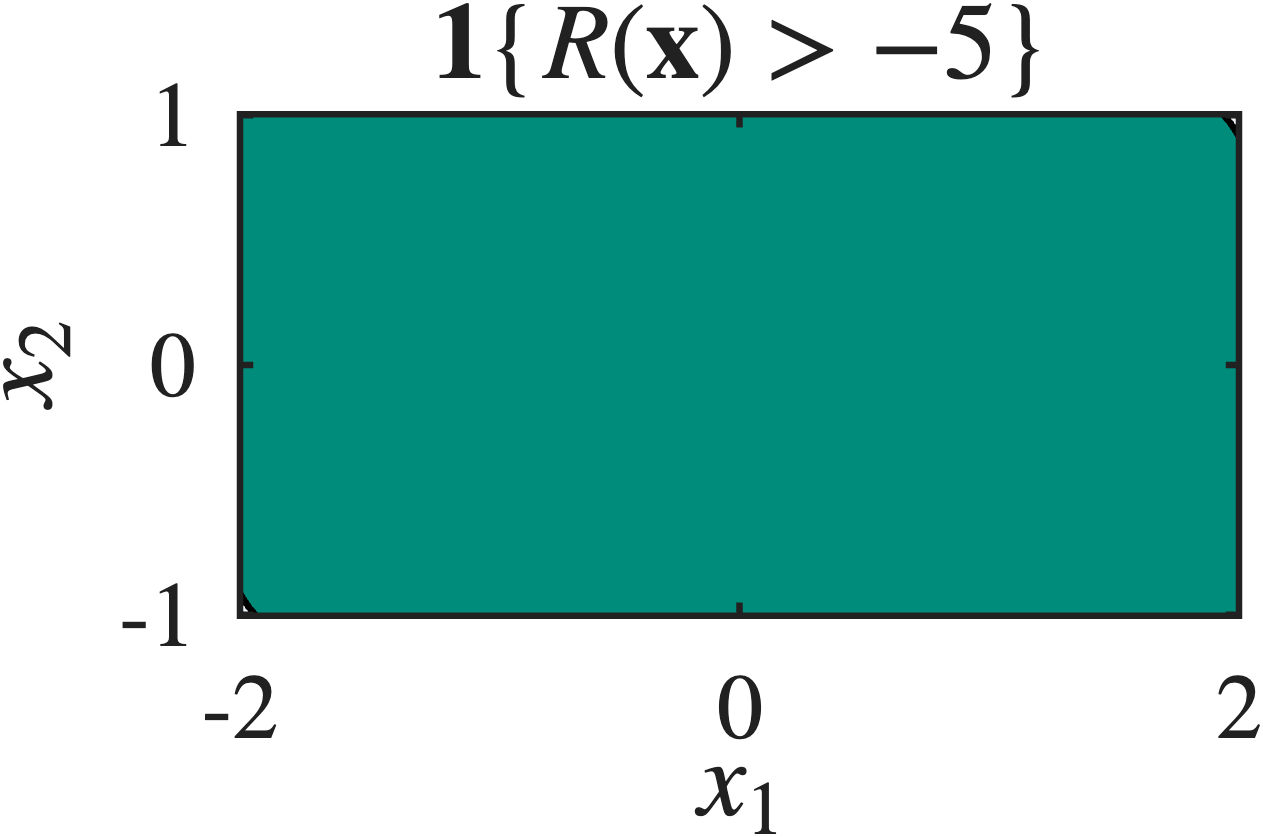}\hfill
\includegraphics[width=.19\linewidth]{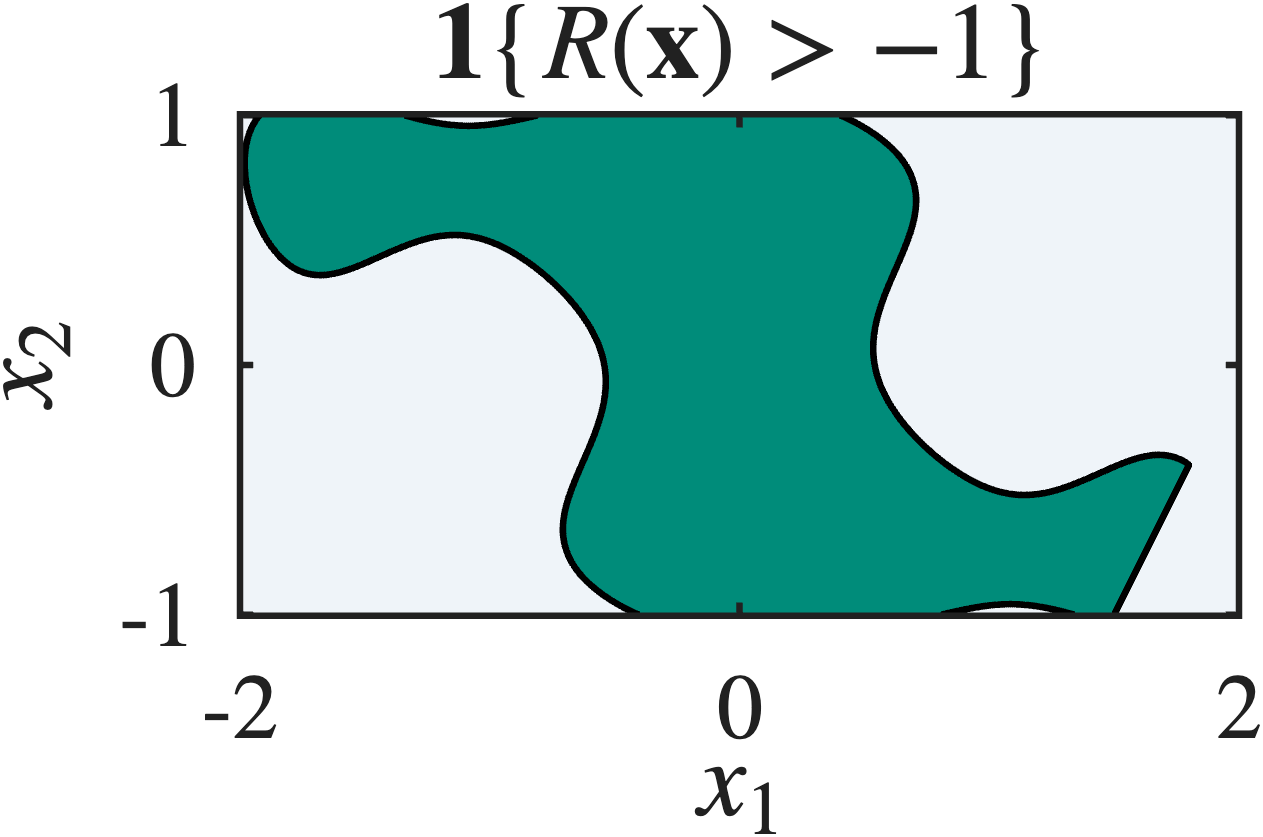}\hfill
\includegraphics[width=.19\linewidth]{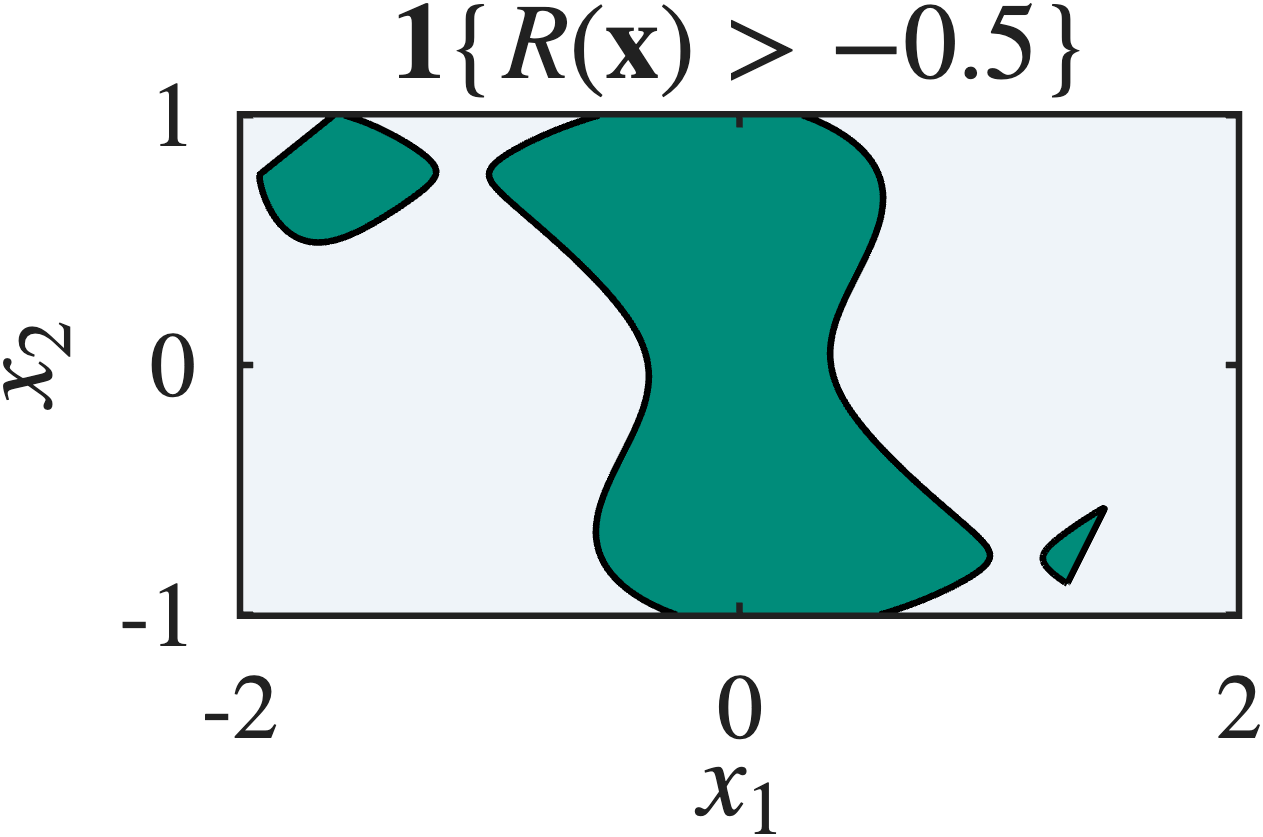}\hfill
\includegraphics[width=.19\linewidth]{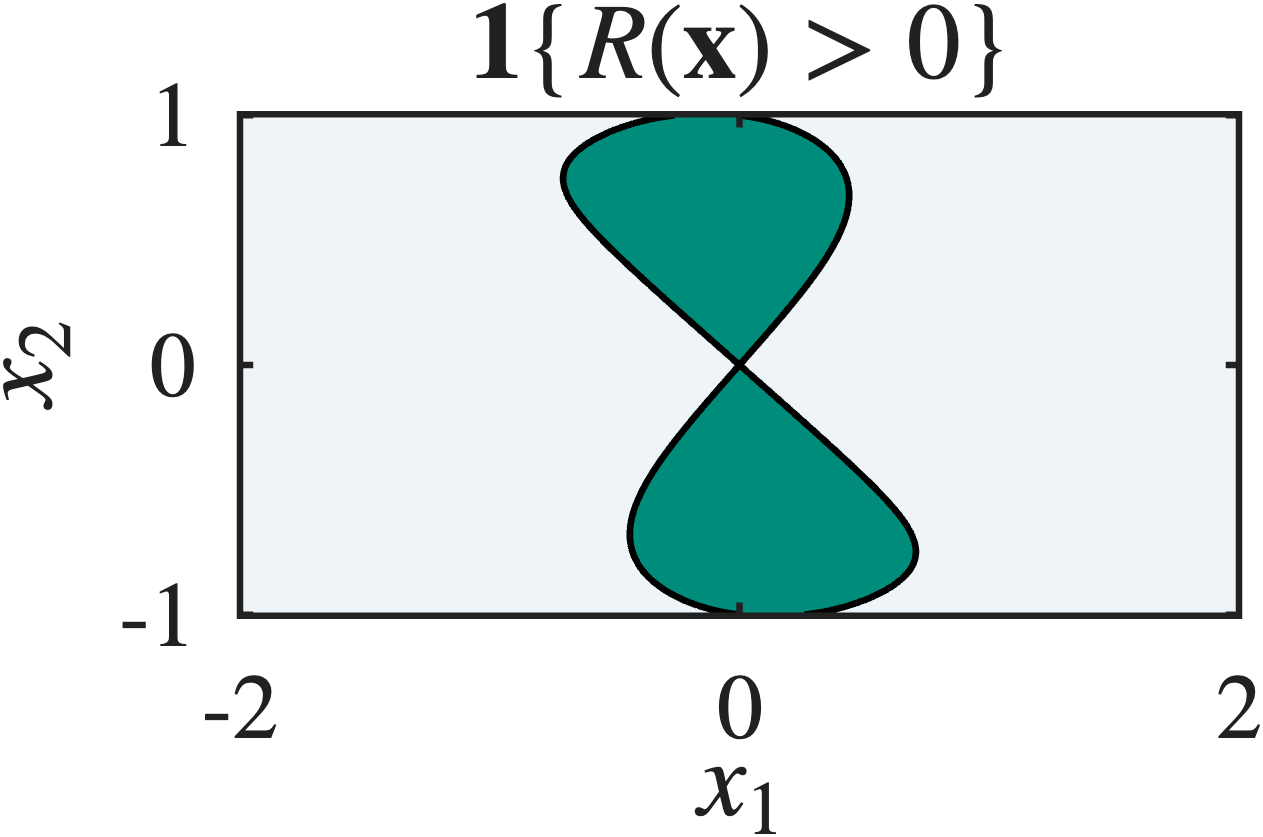}
}
\centerline{
\includegraphics[width=.19\linewidth]{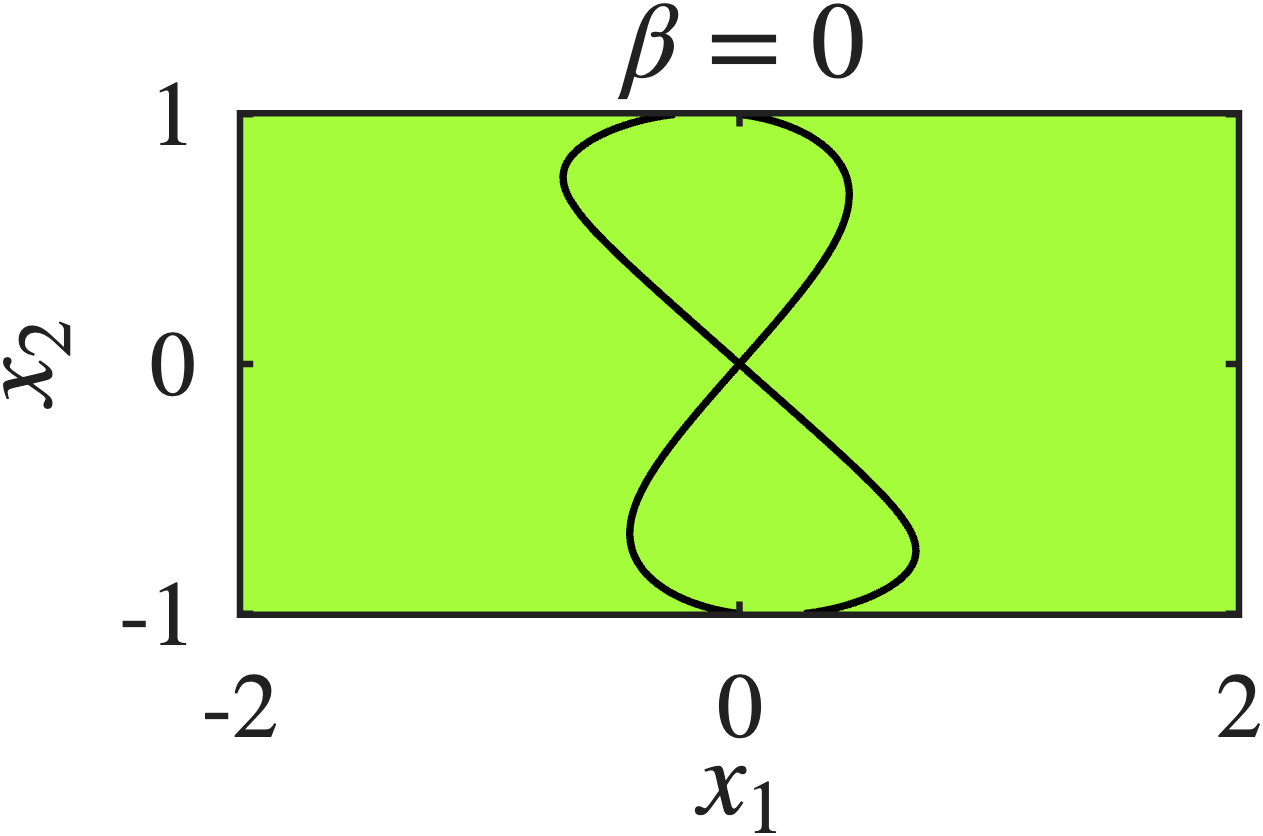}\hfill
\includegraphics[width=.19\linewidth]{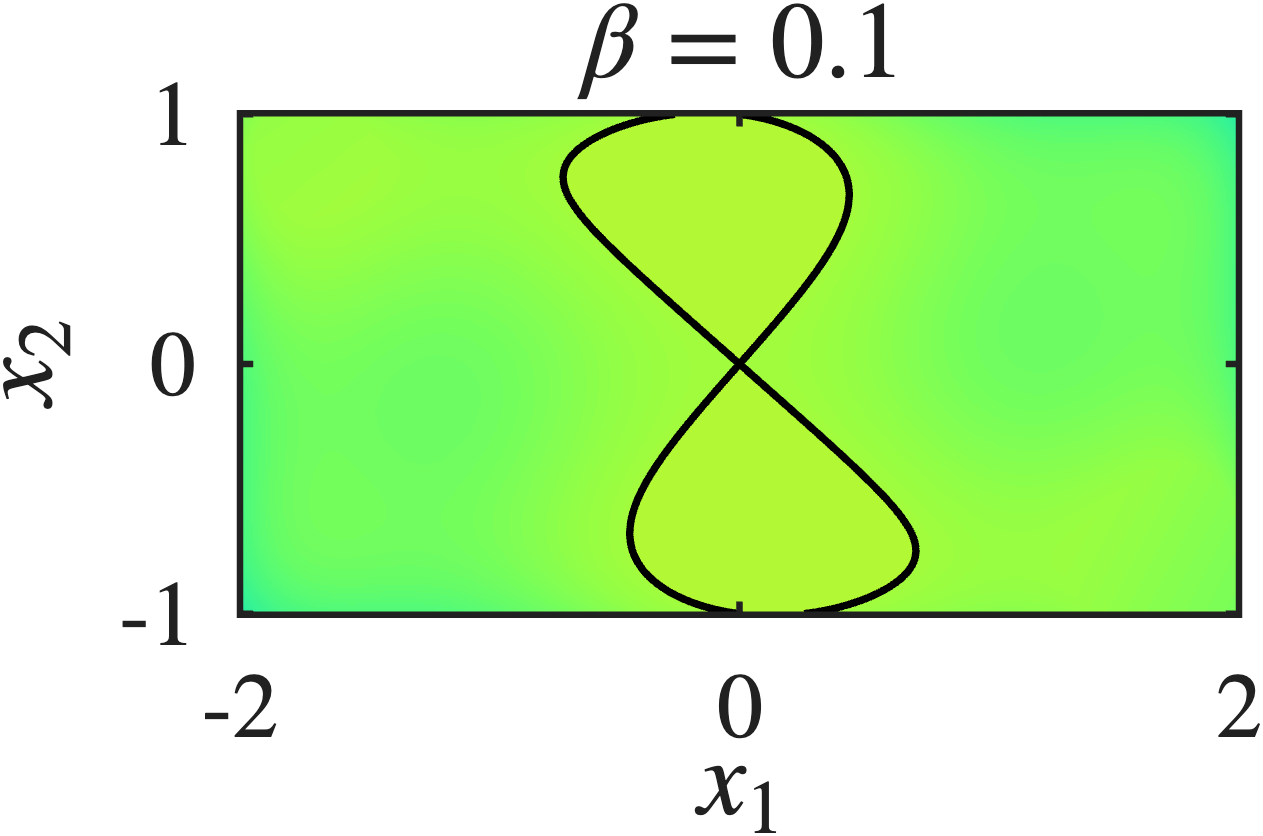}\hfill
\includegraphics[width=.19\linewidth]{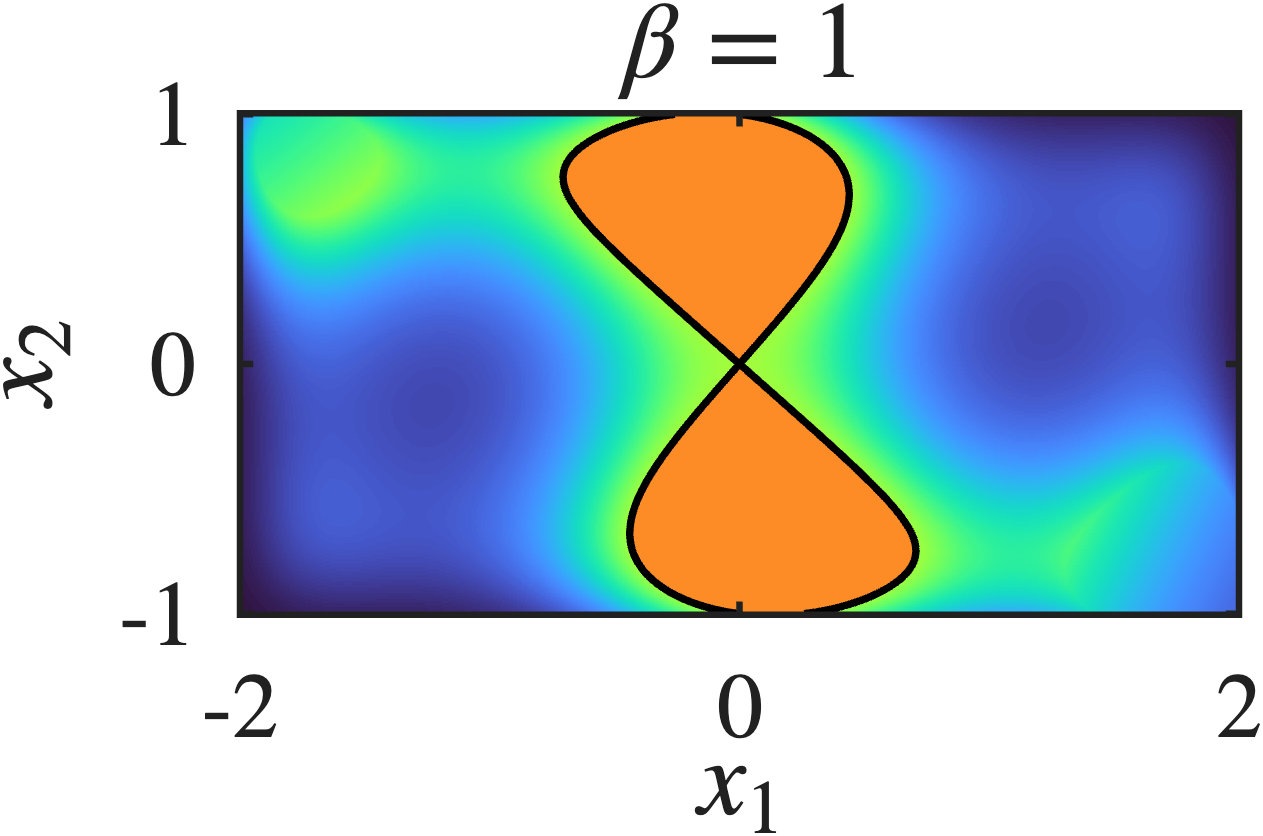}\hfill
\includegraphics[width=.19\linewidth]{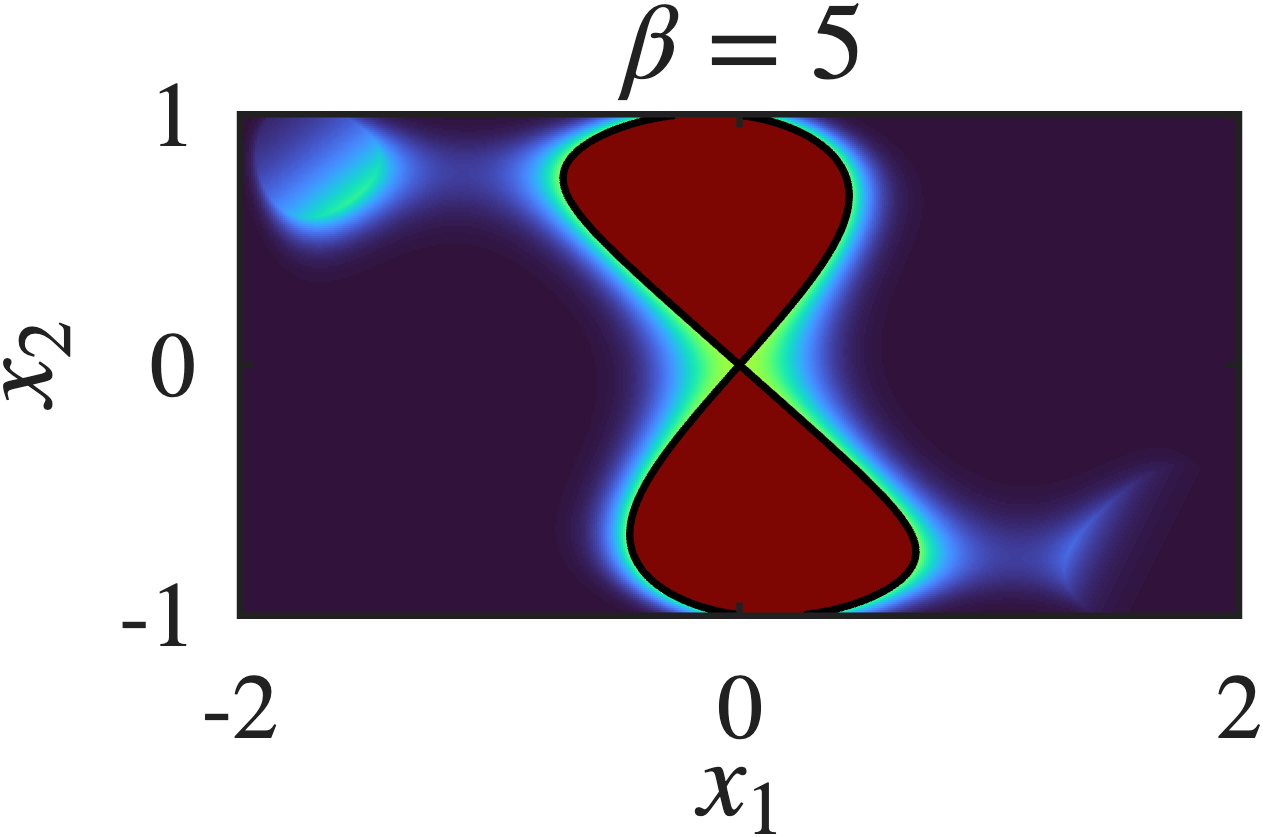}\hfill
\includegraphics[width=.19\linewidth]{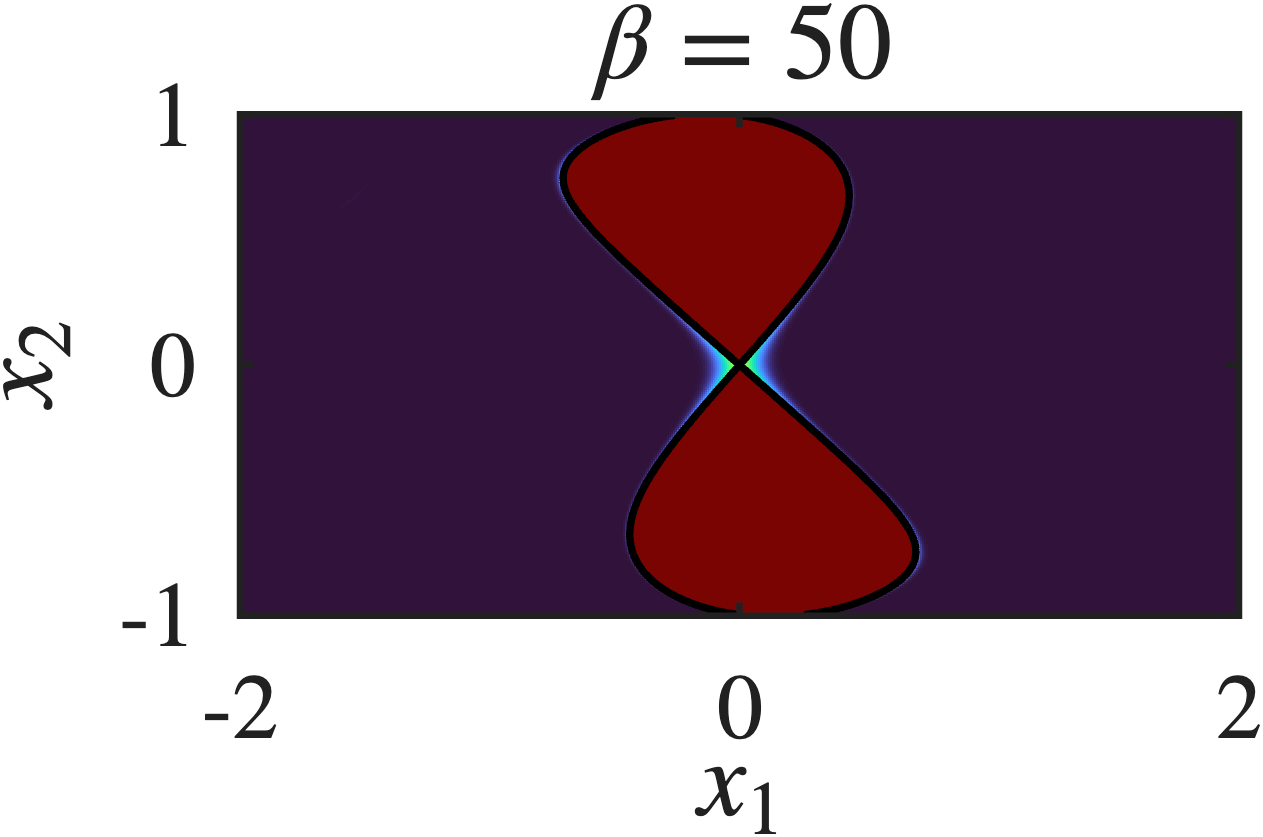}
}
\caption{Top: Hard threshold regions with respect to the $R$-function of the two-dimensional disjoint region of Example~3 in \citet{kucherenko2025rfunctions}. Bottom: Unnormalised logistic intermediate distributions in \eqref{eq:smc-logistic-targets} at increasing inverse temperatures $\beta=0,0.1,1,5,50$ (left to right). Continuous negative R-function values provide graded information outside the feasible region during tempering, while the limiting hard target is uniform on the feasible set.}
\label{fig:bridge}
\end{figure}

\begin{example}[\textbf{SMC for the R-function in Example \ref{ex: Rfunctionexample}}]
We illustrate how continuous R-function information is exploited with SMC. Figure~\ref{fig:bridge} illustrates the intermediate distributions of the SMC sampler applied to the R-function defined in Example \ref{ex: Rfunctionexample}. The top row shows the regions where the $R$-function exceeds an increasing sequence of thresholds from left to right. The rightmost threshold, zero, gives the feasible set, whereas negative thresholds identify progressively wider neighbourhoods of it. This graded information outside the feasible region is used to construct softened intermediate targets in \eqref{eq:smc-logistic-targets} for the SMC sampler. The bottom row in Figure~\ref{fig:bridge} illustrates how the logistic intermediate targets progressively concentrate probability mass around the feasible region as the inverse temperature increases. 
\end{example}

\subsection{Region-only SMC samplers} \label{sec: Region-only annealing with SMC}
To isolate the contribution of the continuous information provided by the R-function, we compare the proposed method with a region-only (RO) construction of intermediate distributions. This construction uses only the binary feasible/infeasible membership, while discarding the magnitude of $R(\mathbf{x})$ outside the feasible region. Given $0 \leq \beta \leq 1$, the unnormalised region-only intermediate distribution is 
\begin{equation} \label{eq:ro-target}
\gamma_\beta^{\rm RO}(\mathbf{x}) = \mathbf 1_{\mathcal R}(\mathbf{x})  + (1-\beta) 
\mathbf 1_{\mathcal A \setminus \mathcal{R}}(\mathbf{x}),
\end{equation}
Its normalising constant is
\begin{equation*}
Z_\beta^{\rm RO} = V+(1-\beta)(A-V) = (1-\beta)A+\beta V.
\end{equation*}
At $\beta=0$, $\gamma_0^{\rm RO}(\mathbf{x})=\mathbf 1_{\mathcal A}(\mathbf{x})$, so the normalised target is uniform on the bounding domain. Increasing $\beta$ progressively suppresses the infeasible region while leaving all feasible points at unit weight. At $\beta=1$, we have  $\gamma_1^{\rm RO}(\mathbf{x})=\mathbf 1_{\mathcal R}(\mathbf{x})$, and the normalised target is therefore exactly uniform on the feasible region.

The important distinction from the R-function-informed SMC construction is the information retained outside $\mathcal R$. For any two infeasible points
$\mathbf{x}_1,\mathbf{x}_2\in\mathcal A\setminus\mathcal R$,
\begin{equation*}
\gamma_\beta^{\rm RO}(\mathbf{x}_1) = \gamma_\beta^{\rm RO}(\mathbf{x}_2) = 1-\beta,
\end{equation*}
irrespective of the values of $R(\mathbf{x}_1)$ and $R(\mathbf{x}_2)$. Thus the region-only construction uses only the sign of the R-function. In contrast, the proposed SMC intermediate distribution uses the same sign information to define feasibility while additionally retaining the continuous values of $R(\mathbf{x})$ for $R(\mathbf{x}) < 0$ during the intermediate stages. Infeasible points can therefore be distinguished according to their constraint-violation scores, without interpreting those scores as geometric distances.

\section{Numerical experiments} \label{sec: Numerical experiments}
\paragraph{Target regions.} For numerical experiments, we consider feasible regions constructed from two $d$-dimensional hyperballs. For each hyperball, the positive-inside primitive is
\begin{equation*}
\phi_i(\mathbf{x})=r_i^2-\|\mathbf{x}-c_i\|^2, \qquad i=1,2,
\end{equation*}
so that $\phi_i(\mathbf{x})\geq0$ if and only if $\mathbf{x}$ lies inside the $i$th hyperball.  The centres are chosen as
\[
c_1=(0,\ldots,0), \qquad c_2=(0,\ldots,0,1),
\] 
so that the centres are one unit apart. The corresponding intersection and union are represented, respectively, by the R-conjunction and R-disjunction
\begin{align}
R_{\wedge}(\mathbf{x}) &= \phi_1(\mathbf{x}) \wedge_\alpha \phi_2(\mathbf{x}), \qquad R_{\vee}(\mathbf{x}) =  \phi_1(\mathbf{x}) \vee_\alpha \phi_2(\mathbf{x}) 
\label{eq:hyperball-rfunctions-union}
\end{align}
For $\alpha=1$, these expressions reduce exactly to $R_{\wedge}=\min \left(\phi_1, \phi_2\right)$ and $R_{\vee}=\max \left(\phi_1, \phi_2\right)$.  Thus $R_{\wedge}(\mathbf{x})\geq0$ defines the intersection, whereas $R_{\vee}(\mathbf{x})\geq0$ defines the union.  The union and intersection volumes of those target regions can be derived analytically; the formulas are given in Appendix~\ref{app:ball-volumes}.

\paragraph{Compared methods.}  
The experiments involve the three methods discussed above: 
\begin{enumerate}
\item \textbf{RS/IS}: When rejection sampling is applied to sample from the feasible region, an importance sampling (IS) estimator can be constructed from its proposal draws. Thus, the volume estimator associated with uniform-proposal rejection sampling is labelled IS.
\item \textbf{SMC}: This stands for the R-function-informed SMC sampler in Section \ref{sec: R-function-informed SMC Samplers}, with intermediate distributions defined in \eqref{eq:smc-logistic-targets},
\item \textbf{RO}: This stands for the region-only SMC sampler in Section \ref{sec: Region-only annealing with SMC}, which uses intermediate distributions in \eqref{eq:ro-target} that only use the binary information about the target region. 
\end{enumerate}

\paragraph{Common settings.} For both SMC and RO, we used $K = 16$ SMC stages, starting from the uniform distribution over $\mathcal{A} = [-5, 5]^{d}$. For the SMC sampler targeting the R-function-informed intermediate distributions $\{ \gamma_{k} \}_{k = 0}^{K}$, we used the schedule 
\[
\beta_{0} = 0, \quad \beta_{k} = \begin{cases} \epsilon +(1 - \epsilon) (k-1)/9 & \text{ for } 1 \leq k \leq 10 \\ 2 (k-10) & \text{ for } 11 \leq k \leq 15 \end{cases}, \quad \beta_{16} = \infty.
\]
with $\epsilon = 0.01$. For RO in \eqref{eq:ro-target}, we considered two alternative schedules: 
\begin{enumerate}
\item \textbf{RO-lin}: The first one is set as
\begin{equation*}
\beta_{k}^{\text{lin}} = k/K, \quad k = 0, \ldots, K,
\end{equation*}
which corresponds to a ``linear" type of interpolation between $A$ and $V$, and yields normalising constants $Z_{k} = V k/K + A(1 - k/K)$.
\item \textbf{RO-geo}: The second alternative is an \emph{oracle benchmark:} $\beta_{k}$ is selected such that $Z_{k}/Z_{k-1}$ is constant over $k$. This can be achieved by solving $(1 - \beta_{k}) A + \beta_{k} V  = Z_{k}$ for $\beta_{k}$ using $Z_{k} =A(V/A)^{k/K}$, which yields
\[
\beta_{k}^{\text{geo}} = \frac{(V/A)^{k/K}-1}{V/A-1}, \quad k = 0, \ldots, K.
\]
Note that this schedule, balanced in terms of the ratio of consecutive normalising constants, is not possible in practice since it requires knowing $V$. However, for the feasible regions in our illustrative experiment, $V$ \emph{is} known (the volume of intersection or union of hyperballs, see Appendix~\ref{app:ball-volumes}). Thus, this scheduling with a constant ratio of successive normalising constants is considered an oracle benchmark.
\end{enumerate}
For both SMC algorithms, we use $N=10000$ particles and one MCMC rejuvenation move per SMC stage, which uses an MH kernel with a Gaussian random-walk proposal. The proposal standard deviations for the random walk MCMC moves used in these experiments are $\sigma_q =0.5$ for R-function-informed SMC and $\sigma_q^{\rm RO}=1$ for RO-lin and RO-geo. 

The RS/IS estimator uses the uniform distribution over $\mathcal{A}$ as its proposal distribution.

Finally, for each feasible region, each method is run independently $B = 1000$ times, from which performance statistics are collected.

\paragraph{Comparison metrics.}
The four methods SMC, RO-lin, RO-geo, and IS are systematically compared pairwise in terms of the accuracy of their volume estimators. For two methods A, B define the variance gain of A over B as
\begin{equation} \label{eq: variance gain}
G^{\text{Var}}_{\textrm{A}/\textrm{B}} = \Var(\widehat V_{\rm B}) / \Var(\widehat V_{\rm A}),
\end{equation}
so values greater than one indicate variance reduction by method A relative to method B. However, the variance gain alone does not take the computation times into account. For this reason, we also computed CPU-adjusted gains of one method over the other for the three comparisons against IS. For methods A and B, this gain is calculated as
\begin{equation} \label{eq: CPU-adjusted variance gain}
G^{\text{CPU}}_{\textrm{A}/\textrm{B}} = G^{\text{Var}}_{\textrm{A}/\textrm{B}} \cdot (\mathrm{CPU}_{\rm B} / \mathrm{CPU}_{\rm A}).
\end{equation}
For the feasible regions involved in our experiment, $V$ is known. This reduces the variability in our comparisons by allowing us to use the exact variance calculations for IS; its $\text{Var}(\widehat{V}_{\text{IS}})$ is calculated exactly as in \eqref{eq:is-variance}. For the other methods, we used sample variances, obtained from $B = 1000$ runs, to approximate the variances of their estimators. 

\paragraph{Timing protocol and environment.}
The experiments were run on MATLAB R2025b on a MacBook Pro (2020) with an Apple M1 processor, eight CPU cores, and 16\,GB RAM. SMC timings include initial particle generation, target evaluations, incremental weighting, normalising-constant estimation, systematic resampling, and MCMC rejuvenation. IS timings include proposal generation, feasibility testing, and volume estimation.  

\subsection{Intersection of two hyperballs} \label{sec:hyperball-intersection}
First, we assess the performance of the compared methods for the intersection of hyperballs, across the dimensions $d=2,\ldots,5$. For each dimension, three intersection geometries are considered: equal radii $(r_1=r_2=1)$ and two variants of unequal radii, $(r_1=1, r_2=0.5)$ and $(r_{1} = 1, r_{2} = 0.25)$. The latter two have substantially smaller feasible volumes and therefore provide more demanding tests for the methods.

\begin{figure}[htbp]
\centering
\includegraphics[width=\linewidth]{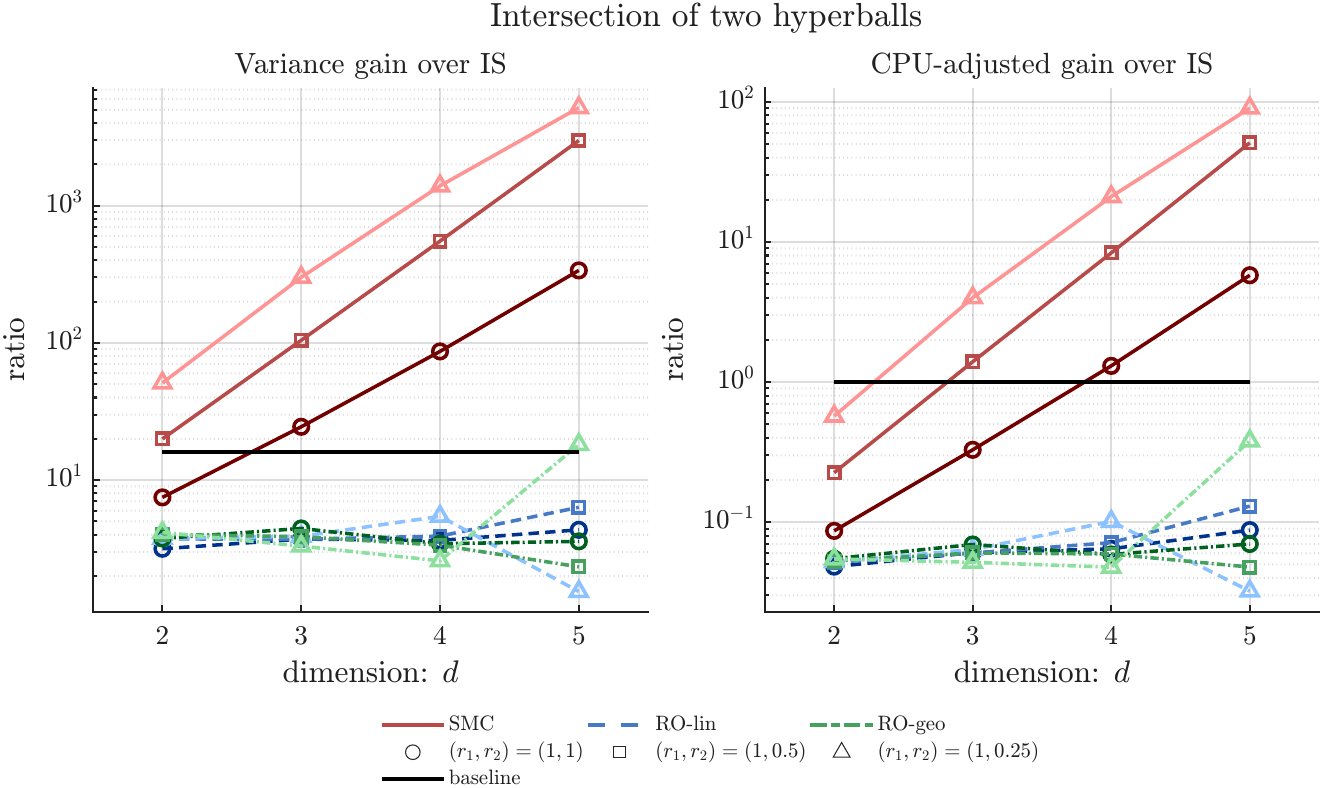}
\caption{Comparison for the intersection experiments.
Left: variance gain over IS, computed as the exact IS variance
$V(A-V)/N$ divided by the empirical variance of SMC, RO-lin, or RO-geo.
Right: the corresponding CPU-adjusted variance gains. SMC, RO-lin, and RO-geo are
shown in red, blue, and green tones, respectively. The black horizontal
lines mark the reference values $K$ on the left and $1$ on the right.}
\label{fig:intersection-combined-mse}
\end{figure}

Figure~\ref{fig:intersection-combined-mse} shows the gains of SMC and the two versions of RO, RO-lin and RO-geo, over the IS estimator in terms of variance and computational efficiency, using the two metrics $G^{\text{Var}}_{\textrm{A}/\textrm{B}}$ and $G^{\text{CPU}}_{\textrm{A}/\textrm{B}}$ defined in \eqref{eq: variance gain} and \eqref{eq: CPU-adjusted variance gain}. The variance gains are compared with the reference value $K=16$, representing a nominal computational-cost scaling based on the number of SMC stages. This is a heuristic comparison rather than an exact equivalence: the costs of target evaluation, weighting, resampling, and MCMC rejuvenation differ from those of independent rejection sampling. The CPU-adjusted gains in the right panel therefore provide a more direct assessment of computational efficiency, with values above $1$ indicating an advantage.


Figure~\ref{fig:intersection-combined-mse} reveals that the variance advantage of SMC increases strongly with dimension and is more pronounced for the smaller unequal-radius intersection. The CPU-adjusted results show the corresponding trade-off: at lower dimensions the variance gain of SMC over IS does not fully compensate for the additional cost of the sequential SMC stages, whereas at higher dimensions SMC becomes advantageous even after computational cost is taken into account. In contrast, both RO-lin and RO-geo are below a CPU-adjusted gain of $1$. This shows the advantage of using R-function-informed intermediate distributions in SMC.

\subsection{Union of two hyperballs}
\label{sec:hyperball-union}

We next consider the R-disjunction $R_{\vee}$ in Eq.~\eqref{eq:hyperball-rfunctions-union}. For each dimension, six union geometries are considered: three variants of equal radii $(r_1 = r_2 \in \{ 0.25, 0.5, 1\})$, and three variants of unequal radii, $( 2 r_2 = r_1 \in \{ 0.25, 0.5, 1\})$. When $r_1 \in \{0.25, 0.5 \}$, the two hyperballs are tangent or disjoint, so their interiors are disconnected. This provides a more challenging rare-event test and, at the same time, tests whether the SMC procedure can maintain particles in separated feasible components.

\begin{figure}[htbp]
\centering
\includegraphics[width=\linewidth]{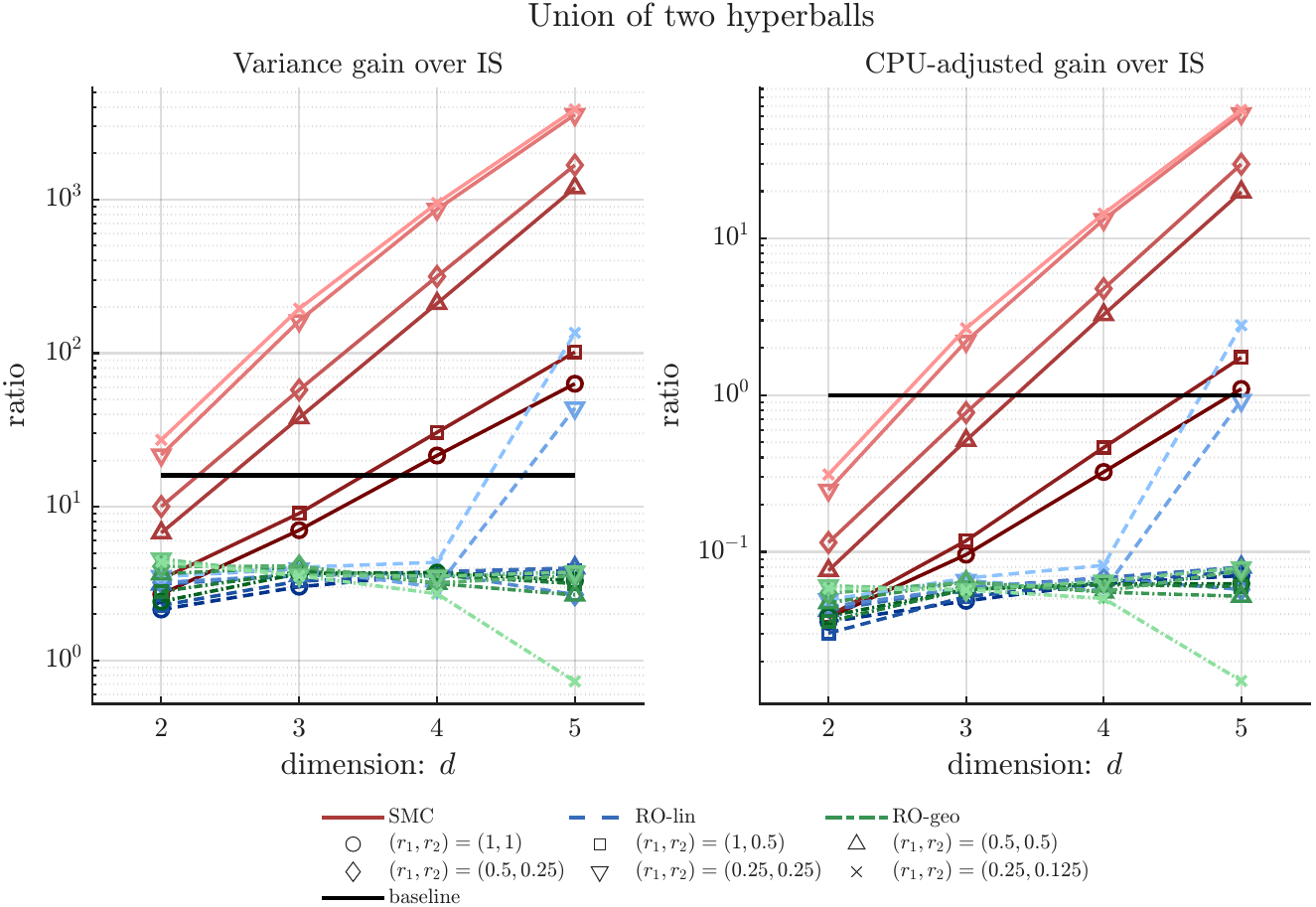}
\caption{Comparison for the union experiments. Left: variance gain over IS, computed as the exact IS variance
$V(A-V)/N$ divided by the empirical variance of SMC, RO-lin, or RO-geo. Right: the corresponding CPU-adjusted variance gains. SMC, RO-lin, and RO-geo are shown in red, blue, and green tones, respectively. The black horizontal lines mark the reference values $K$ on the left and $1$ on the right.}
\label{fig:union-combined-mse}
\end{figure}

Figure~\ref{fig:union-combined-mse} compares SMC, RO-lin and RO-geo against IS using variance and CPU-adjusted variance gains. The SMC advantage generally increases as the feasible fraction decreases. RO-geo remains below a CPU-adjusted gain of one in these experiments, whereas RO-lin exceeds one in one rare union configuration. However, such isolated gains require caution when nearly all volume estimates are zero, as shown in Table~\ref{tbl: zero volume estimates}.

\paragraph{Bootstrap intervals for the gains.}
In addition to the point estimates of the gains, we report the bootstrap intervals for the variance and CPU-adjusted gains in Tables \ref{tab:boot-intersection} and \ref{tab:boot-union} in Appendix \ref{app:bootstrap}. We quantify uncertainty in the gains using 5000 nonparametric bootstrap resamples of the 1000 runs for each configuration. The bootstrap intervals may be useful especially for small regions where the variation is high. For example, the point estimate $2.79$ of the CPU-adjusted gain of RO-lin for the $d=5$ union $(0.25,0.125)$ has a bootstrap interval $[0.926,\infty)$, which indicates that the CPU-adjusted advantage implied by the point estimate $2.79$ is not established and may be due to Monte Carlo variation.

\FloatBarrier
\subsection{Additional results}
The comparisons on the intersection and union cases show the advantage of R-function-informed SMC samplers. The following sections provide further insight into sampling accuracy and particle evolution.

\subsubsection{Sampling accuracy}

\begin{table}[!hbtp]
\centering
\caption{Counts of exactly zero volume estimates among $B=1000$ runs at $d=5$.}\label{tbl: zero volume estimates}
\begin{tabular}{@{}lrrrr@{}}\toprule
Geometry and radii & IS & SMC & RO-lin & RO-geo \\
\midrule
Intersection $(1,0.25)$ & 999 & 0 & 999 & 996 \\
Union $(0.25,0.25)$ & 999 & 0 & 997 & 986 \\
Union $(0.25,0.125)$ & 1000 & 0 & 999 & 992 \\
\bottomrule
\end{tabular}
\end{table}

\begin{figure}[!htbp]
\centering
\includegraphics[width=\linewidth]{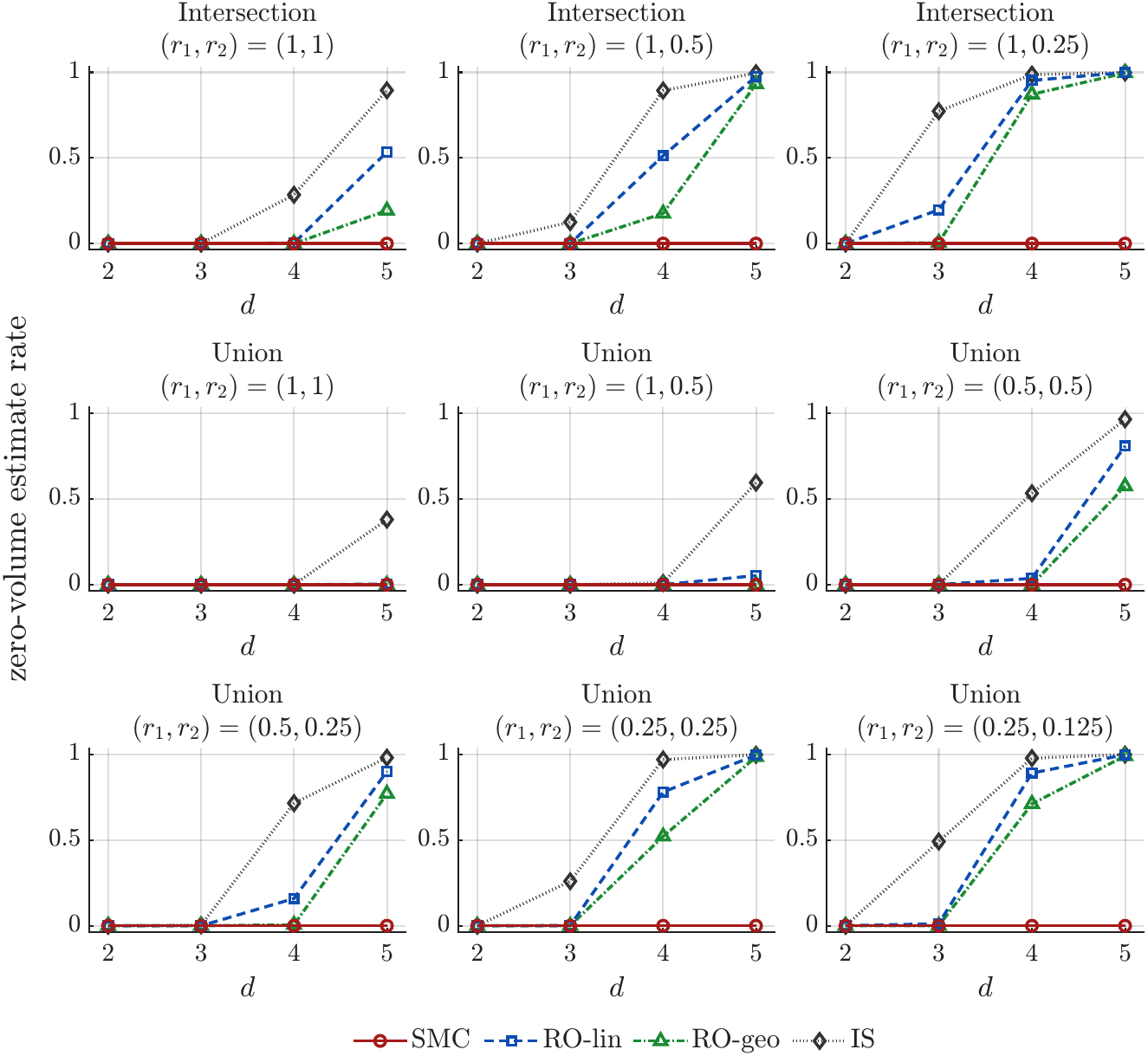}
\caption{Rates of zero-volume estimates, calculated from $B = 1000$ Monte Carlo runs for each model.}
\label{fig:zero-volume estimators}
\end{figure}

\paragraph{Zero-volume estimates.}
Table~\ref{tbl: zero volume estimates} reports the counts of zero volume estimates among $B=1000$ runs at $d=5$ for three rare configurations. SMC has no zero estimates in these configurations. IS returns zero in 999, 999, and 1000 runs, respectively, while both region-only methods also return zero in most runs. Figure \ref{fig:zero-volume estimators} gives a more complete summary by showing the zero-estimate rates of all the methods for all geometries. These results support the robustness of SMC in the reported configurations; the rare nonzero outcomes should also be considered when interpreting empirical variance gains.

\FloatBarrier
\paragraph{Component occupancy.}
In addition to the rate of zero-volume estimates, we also assess the sampling accuracy of the methods. For this, we consider the rate of occupancy in one of the hyperballs for unions of two hyperballs with non-overlapping interiors, i.e., those with $r_1+r_2\leq\|c_1-c_2\|=1$. The true probability of the first ball under the uniform distribution on the union and its empirical occupancy fraction are
\[
p_1=\frac{v_d(r_1)}{v_d(r_1)+v_d(r_2)} = \frac{r_1^d}{r_1^d+r_2^d}, \qquad \widehat p_1=\frac{n_1}{n_1+n_2},
\]
where $v_d(r)$ is the volume of a $d$-dimensional ball with radius $r$. To assess the accuracy of the particle approximation, we compare the particle estimate of $p_{1}$ to its true value. A particle estimate of $p_{1}$ is available only when a run is successfully completed, i.e., with at least one particle with non-zero weight at $k = K$. For a successfully completed run, let $n_i$ count final post-resampling, post-rejuvenation particles in ball $i$, so that $n_1+n_2=N$. 
 A failed run, in which the sampler has zero total terminal importance weight and hence no valid final feasible population, has no occupancy estimate and is excluded from the boxplot.

\begin{figure}[!htbp]
\centering
\includegraphics[width=0.49\linewidth]{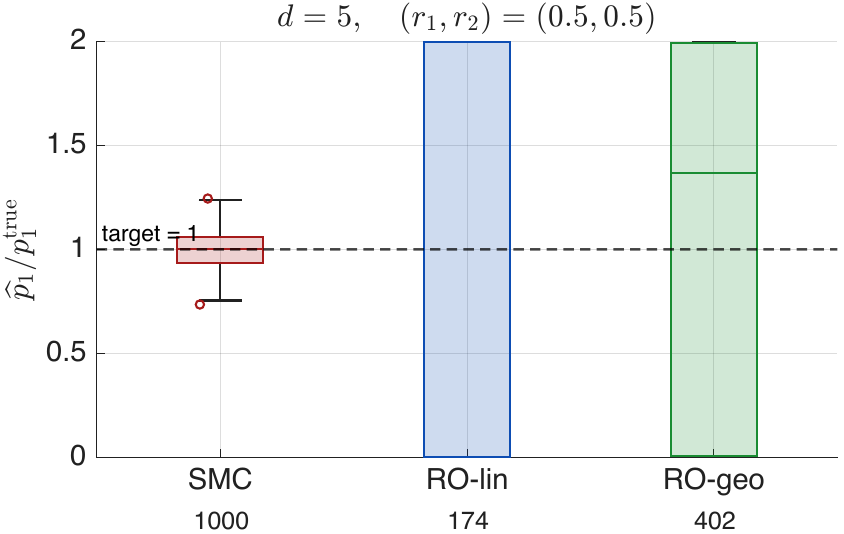}\hfill
\includegraphics[width=0.49\linewidth]{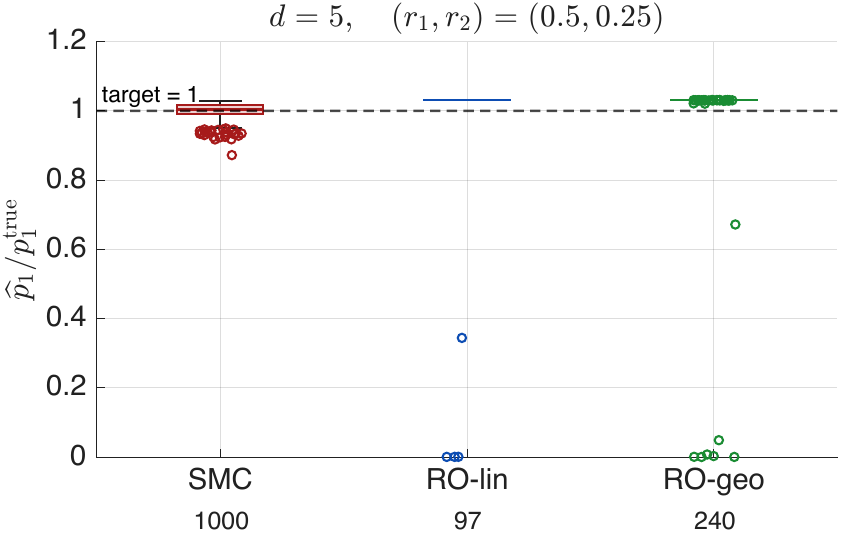}
\par\smallskip
\includegraphics[width=0.49\linewidth]{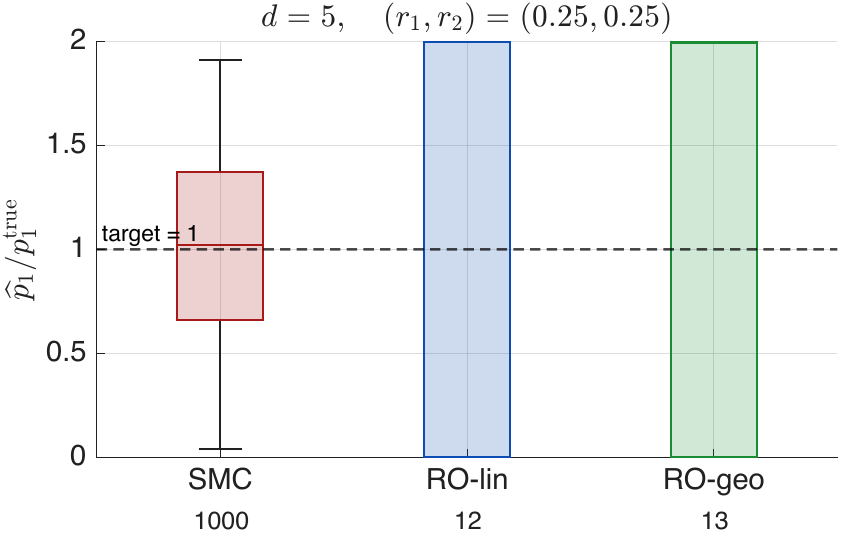}\hfill
\includegraphics[width=0.49\linewidth]{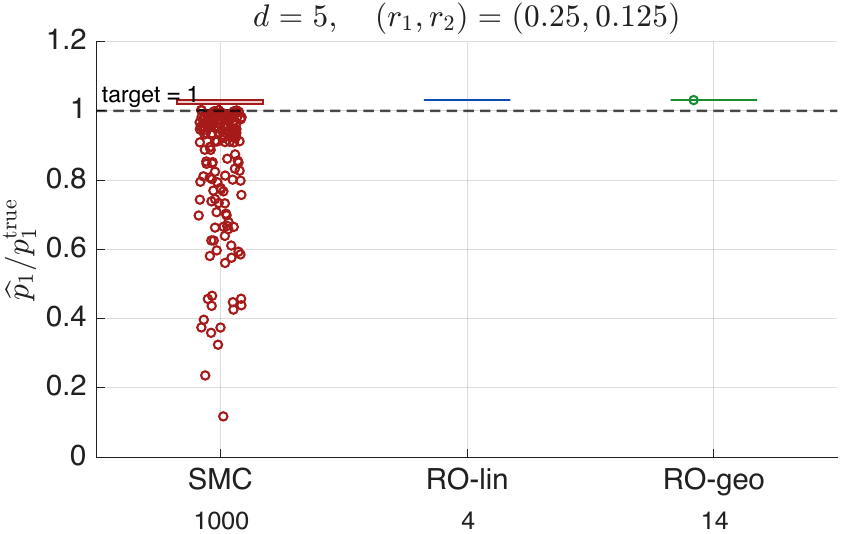}
\caption{Normalised occupancy of the first ball, $Q_1=\widehat p_1/p_1$, for the four zero-overlap union configurations at $d=5$. Boxes include all successful runs (numbers indicated below), which end up with at least one particle inside the feasible region, out of $B = 1000$ runs. Boxes indicate the interquartile range and median; whiskers extend to the most extreme observations within $1.5$ interquartile ranges of the box. Outliers are jittered horizontally to reveal repeated values, without changing their vertical coordinates.}
\label{fig:component-occupancy-d5}
\end{figure}

Figure~\ref{fig:component-occupancy-d5} shows $Q_1=\widehat p_1/p_1$ at $d=5$, with $N=10000$ particles and $1000$ independent runs per method and radius pair. The number of successful runs for each method is indicated below the corresponding label on the x-axis. The results show that the component occupancy estimates from SMC are more reliable than those from RO-lin and RO-geo. Note that RO-lin and RO-geo successfully complete far fewer runs for small $(r_{1}, r_{2})$ configurations. Therefore, a narrow box based on a small surviving subset should not be interpreted as evidence of reliable unconditional performance.

\FloatBarrier
\subsubsection{Effect of R-function information on particle evolution}
\label{sec:particle-evolution}

\begin{figure}[p]
\centering
\includegraphics[width=0.33\textwidth]{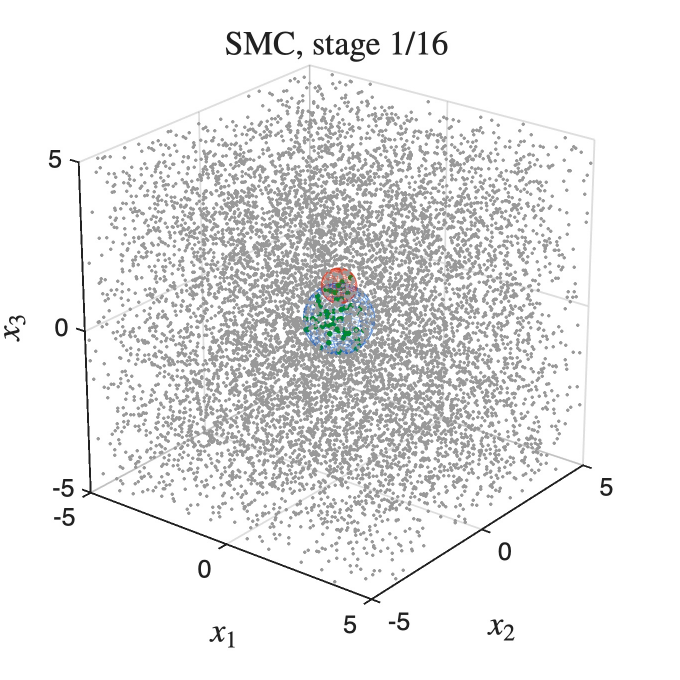}\hfill
\includegraphics[width=0.33\textwidth]{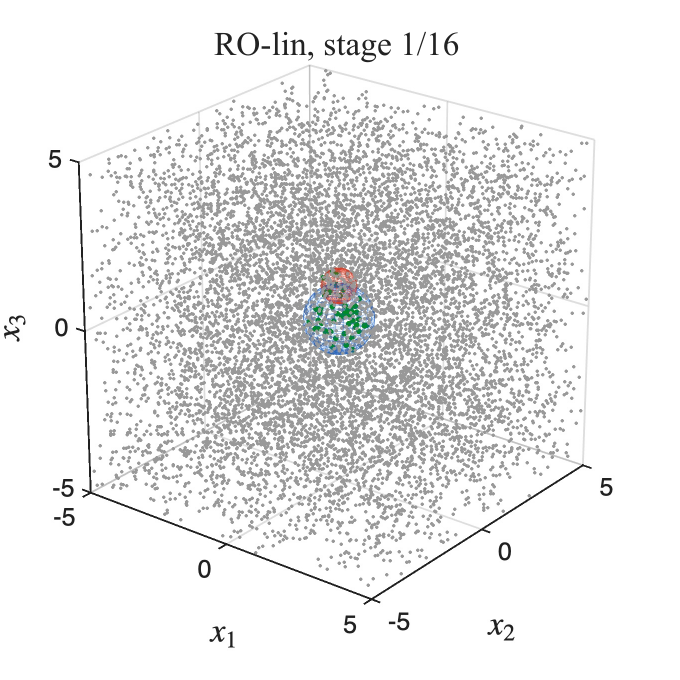}
\par\smallskip
\includegraphics[width=0.33\textwidth]{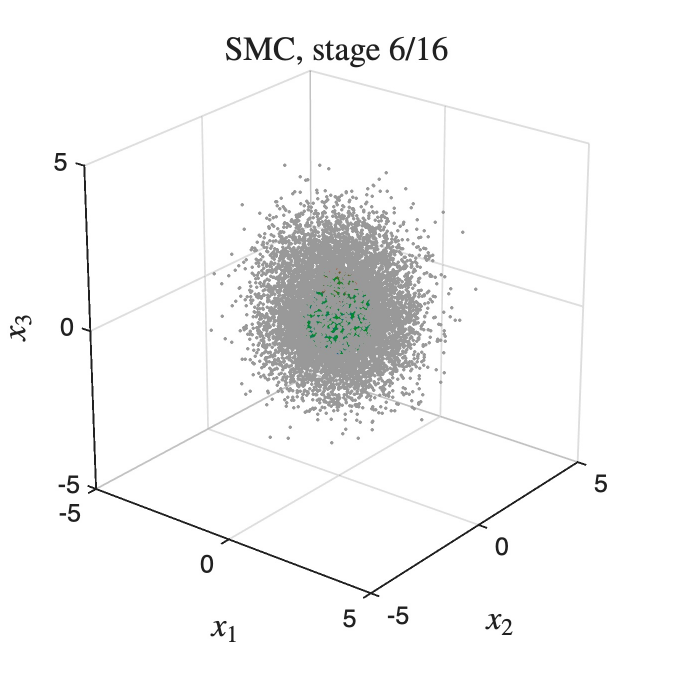}\hfill
\includegraphics[width=0.33\textwidth]{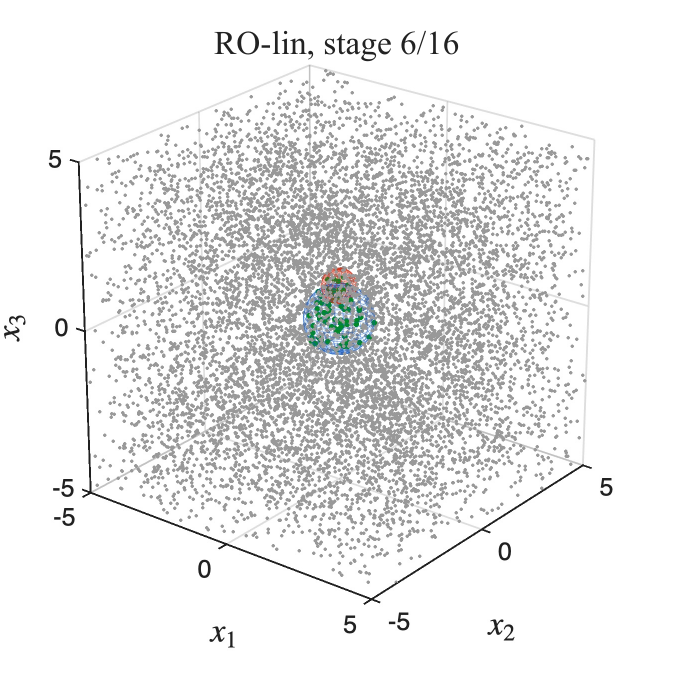}
\par\smallskip
\includegraphics[width=0.33\textwidth]{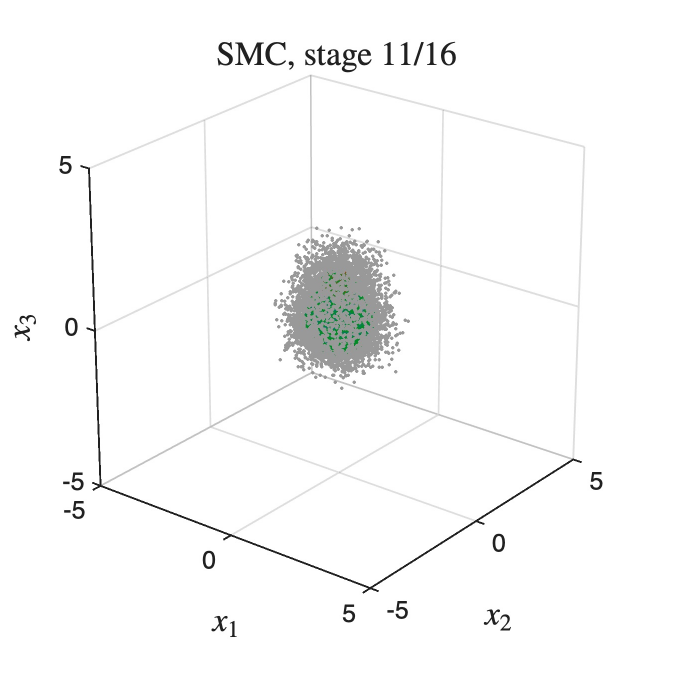}\hfill
\includegraphics[width=0.33\textwidth]{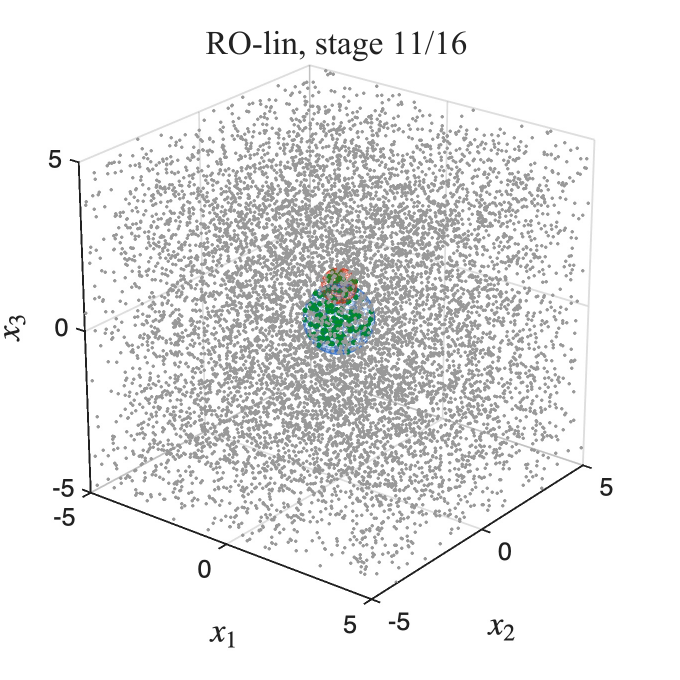}
\par\smallskip
\includegraphics[width=0.33\textwidth]{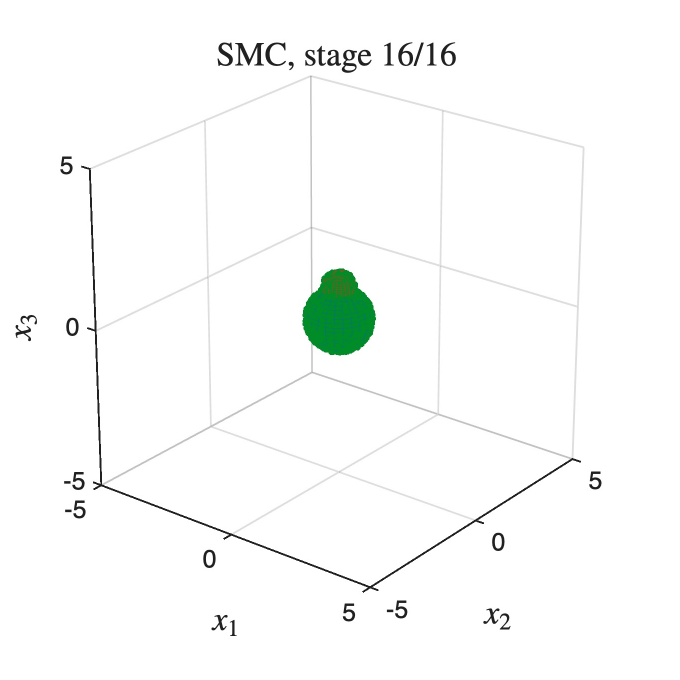}\hfill
\includegraphics[width=0.33\textwidth]{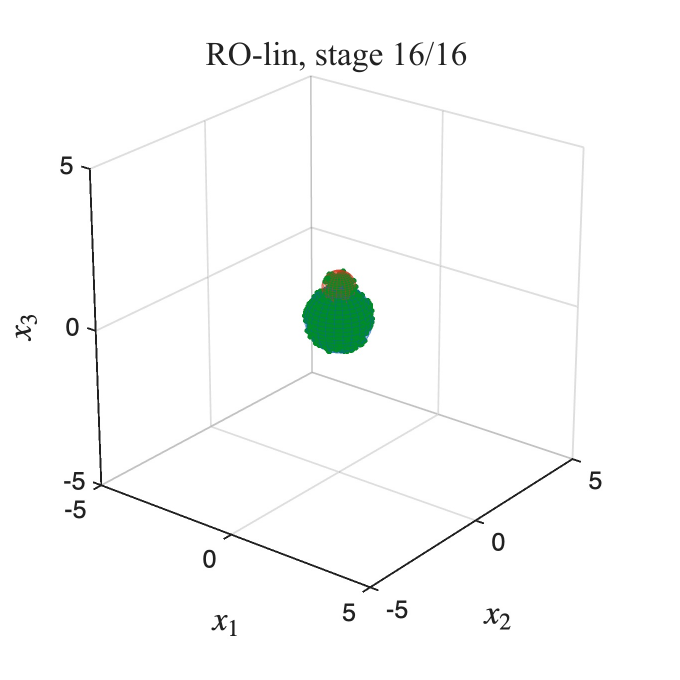}
\caption{Evolution of the particle populations for the three-dimensional union with $(r_1,r_2)=(1,0.5)$ for a single run. Rows correspond to stages $1$, $6$, $11$, and $16$; the left column is the R-function-informed SMC sampler, and the right column is RO-lin. Green points are feasible, and grey points are infeasible.}
\label{fig:smc-sa-evolution}
\end{figure}

\begin{figure}[!htbp]
\centering
\includegraphics[width=0.92\textwidth]
{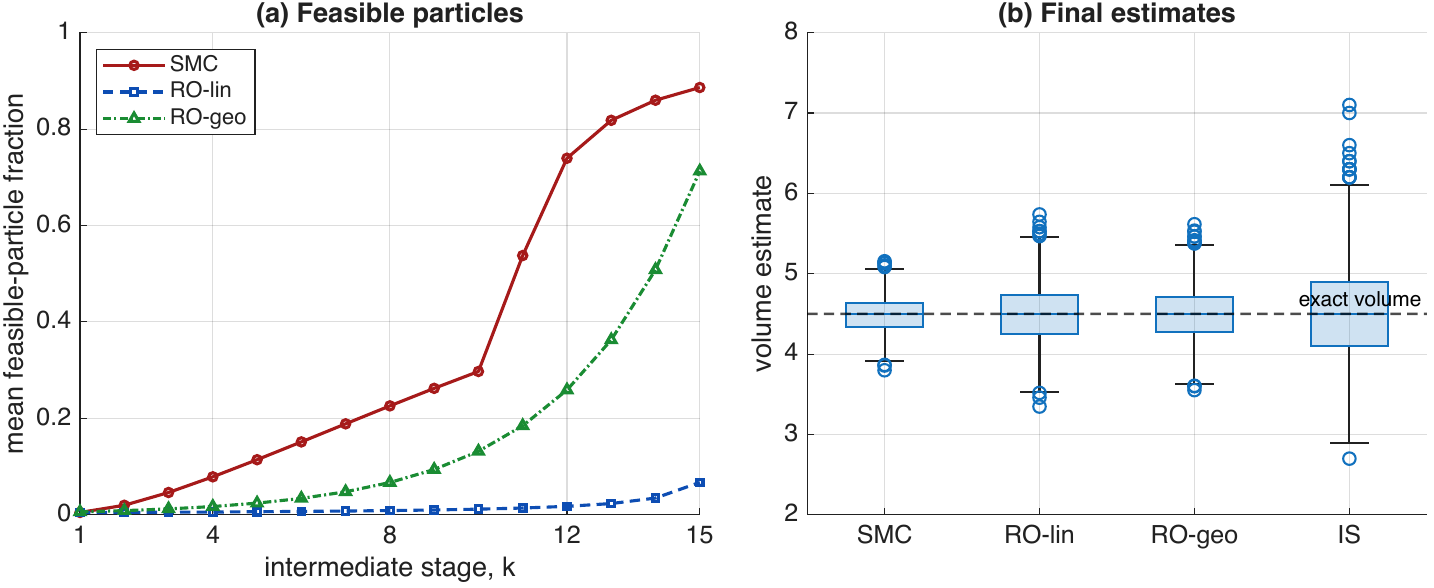}
\caption{Diagnostics for the three-dimensional union with $(r_1,r_2)=(1,0.5)$ over $1000$ independent replications. Left: mean feasible-particle fraction after resampling and rejuvenation for SMC, RO-lin and RO-geo at stages $1$--$15$. Stage $K=16$ is omitted because, conditional on successful hard-target resampling, the fraction is one by construction. Right: final volume estimates for SMC, RO-lin, RO-geo and IS; the dashed line is the exact volume.}
\label{fig:current-diagnostics}
\end{figure}

Figure~\ref{fig:smc-sa-evolution} illustrates the particle evolution for SMC and RO-lin for the three-dimensional union with $(r_1,r_2)=(1,0.5)$. The rows correspond to four intermediate steps, while the left and right columns correspond to the R-function-informed SMC and RO-lin, respectively. Each panel is a three-dimensional scatterplot in $(x_1,x_2,x_3)$; grey points are infeasible, green points are feasible. The figure illustrates the marked difference in how effectively SMC and RO-lin concentrate particles near the feasible region.

Figure~\ref{fig:current-diagnostics} summarises all $B=1000$ replications for the three-dimensional union with $(r_1,r_2)=(1,0.5)$. The left panel shows that all three methods concentrate towards feasibility, but at different rates under their method-specific schedules. The right panel summarises final volume estimates. The R-function-informed SMC sampler concentrates particles in the feasible region earlier and produces less variable volume estimates than the region-only alternatives in this configuration. The concentration curves describe the method-specific schedules; they do not by themselves establish estimator accuracy or computational efficiency. 

\FloatBarrier
\section{Conclusions} \label{sec: Conclusions}

Continuous signed feasibility information can be incorporated directly into an SMC bridge, allowing particles to be guided towards a feasible set before the final hard indicator target is imposed. Across the examples considered here, the R-function-informed SMC approach produces accurate volume estimates and substantial variance reductions relative to direct rejection sampling, with the advantage becoming strongest as the feasible fraction of the bounding domain decreases, for example with increasing dimension in these benchmarks.

To distinguish the effect of continuous R-function information from that of sequential tempering itself, the proposed method was also compared with a region-only (RO) SMC construction. RO uses the same sequential particle machinery but retains only binary feasible/infeasible information, assigning the same finite-temperature weight to all infeasible points. Comparison with RO with both linear and geometric annealing schedules shows earlier concentration in the feasible region and lower dispersion of the final volume estimates for the R-function-informed SMC. These comparisons support the practical value of the chosen bridge, which uses the graded information contained in the continuous negative values of the R-function.  However, differences in schedules prevent attributing the entire gain solely to the graded information provided by the R-function.

Extension of the framework to constrained optimisation is the subject of separate work. Constrained optimisation with R-function-informed SMC samplers can be carried out by involving the objective function $f(\mathbf{x})$ (to be maximised) in the intermediate distributions, such as $\gamma_{k}^{\text{opt}}(\mathbf{x}) = \gamma_{k}(\mathbf{x}) \exp(\beta_{k}^{\text{opt}}f(\mathbf{x}))$ for some increasing sequence $\beta_{k}^{\text{opt}}$, $k = 0, \ldots, K$.

The methodology, along with its implementation, can potentially be improved in several ways. First, we considered random-walk MH for the MCMC move. Gradient-based moves such as Hamiltonian Monte Carlo (HMC, also known as hybrid Monte Carlo) \citep{duane1987hybrid, neal2011mcmc} or the Metropolis-adjusted Langevin algorithm (MALA) \citep{roberts1996exponential} can be employed instead of random-walk MH to improve the rejuvenation step. Relatedly, differentiable intermediate distributions can be considered as an alternative to our choice in \eqref{eq:smc-logistic-targets}. Moreover, we only investigated the choice $\alpha = 1$ to construct R-conjunction and R-disjunction operators; the effect of $\alpha \in (-1, 1)$ could be investigated theoretically and systematically to improve performance.

Overall, the results demonstrate that continuous information about constraint violation can substantially improve the statistical efficiency of SMC-based feasible-region sampling, particularly when the feasible region occupies a small fraction of the bounding domain. CPU-adjusted performance depends on the experimental setting.

\section*{Acknowledgements}
S.K. gratefully acknowledges Prof. Nilay Shah of Imperial College London
for his interest in and support of this work.


\bibliographystyle{apalike}
\bibliography{references}

\newpage
\appendix
\captionsetup{font={footnotesize}}
\section{Analytic volumes of two balls}\label{app:ball-volumes}
Let $B_i=\{x\in\mathbb R^d:\|x-c_i\|\leq r_i\}$, with $r_i>0$, and put
$\delta=\|c_1-c_2\|$. The volume of a ball is
\[
v_d(r)=\frac{\pi^{d/2}r^d}{\Gamma(d/2+1)}.
\]
The union and intersection volumes are given by
\begin{align*}
|B_1\cup B_2| &=v_d(r_1)+v_d(r_2)-|B_1\cap B_2|, \\
|B_1\cap B_2| &=
\begin{cases}
0,&\delta\geq r_1+r_2,\\
v_d (\min\{r_1,r_2\}),&\delta\leq|r_1-r_2|,\\
C_d(r_1,h_1)+C_d(r_2,h_2),&|r_1-r_2|<\delta<r_1+r_2
\end{cases}
\end{align*}
where, in the partial-overlap case,
\[
a_1=\frac{\delta^2+r_1^2-r_2^2}{2\delta},\quad
a_2=\delta-a_1,\qquad h_i=r_i-a_i.
\]
Here $C_d(r,h)$ is a spherical cap's volume, with cap height $h\in[0,2r]$.
Cross-sectional integration gives
\[
C_d(r,h)=v_{d-1}(1)\int_{r-h}^{r}(r^2-t^2)^{(d-1)/2}\,dt.
\]
For $0\leq h\leq r$, the substitution $u=1-(t/r)^2$ expresses the integral in terms of the regularised incomplete beta function; for larger caps, symmetry gives the complementary volume:
\[
C_d(r,h)= \begin{cases}\frac{v_d(r)}{2}
 I_{1-(1-h/r)^2}\left(\frac{d+1}{2},\frac12\right), & \text{ for } 0\leq h\leq r, \\
 v_d(r)-C_d(r,2r-h) &  \text{ for } r<h\leq2r. \end{cases}
\]
where, for $a,b>0$ and $0\leq z\leq1$, the regularised incomplete beta function $I_z(a,b)$ is
\[
I_z(a,b)=\frac{B_z(a,b)}{B(a,b)},\quad
B_z(a,b)=\int_0^z t^{a-1}(1-t)^{b-1}\,dt,\quad
B(a,b)=\frac{\Gamma(a)\Gamma(b)}{\Gamma(a+b)}.
\]

\section{Bootstrap uncertainty for the reported gains} \label{app:bootstrap}
The following tables supplement the point estimates in Figures \ref{fig:intersection-combined-mse}--\ref{fig:union-combined-mse} using the sample-variance and unpaired run-resampling procedure described in the numerical section. For sorted bootstrap values $G^*_{(1)},\ldots,G^*_{(5000)}$, the interval is $[G^*_{(125)},G^*_{(4875)}]$.

\subsection{Intersection experiments}
{\small
\begin{longtable}{@{}clrrr@{}}
\caption{Point estimates and pointwise 95\% percentile bootstrap intervals for intersection variance gains. $z^*$ is the percentage of zero-variance bootstrap resamples.}\label{tab:boot-intersection}\\
\toprule
$d$ & Method & $G^{\rm Var}$ [95\% interval] & $G^{\rm CPU}$ [95\% interval] & $z^*$ (\%) \\
\midrule\endfirsthead
\toprule
$d$ & Method & $G^{\rm Var}$ [95\% interval] & $G^{\rm CPU}$ [95\% interval] & $z^*$ (\%) \\
\midrule\endhead
\multicolumn{5}{l}{$(r_1,r_2)=(1,1)$}\\
2 & SMC & $\num{7.49}$ [$\num{6.86}$, $\num{8.19}$] & $\num{0.0866}$ [$\num{0.0753}$, $\num{0.0985}$] & 0.0 \\
2 & RO-lin & $\num{3.16}$ [$\num{2.90}$, $\num{3.47}$] & $\num{0.0481}$ [$\num{0.0441}$, $\num{0.0530}$] & 0.0 \\
2 & RO-geo & $\num{3.76}$ [$\num{3.46}$, $\num{4.11}$] & $\num{0.0548}$ [$\num{0.0504}$, $\num{0.0600}$] & 0.0 \\
3 & SMC & $\num{24.5}$ [$\num{22.6}$, $\num{26.7}$] & $\num{0.328}$ [$\num{0.303}$, $\num{0.359}$] & 0.0 \\
3 & RO-lin & $\num{3.72}$ [$\num{3.41}$, $\num{4.09}$] & $\num{0.0606}$ [$\num{0.0555}$, $\num{0.0666}$] & 0.0 \\
3 & RO-geo & $\num{4.44}$ [$\num{4.07}$, $\num{4.89}$] & $\num{0.0692}$ [$\num{0.0634}$, $\num{0.0762}$] & 0.0 \\
4 & SMC & $\num{86.7}$ [$\num{79.8}$, $\num{94.6}$] & $\num{1.30}$ [$\num{1.20}$, $\num{1.42}$] & 0.0 \\
4 & RO-lin & $\num{3.64}$ [$\num{3.34}$, $\num{3.97}$] & $\num{0.0645}$ [$\num{0.0593}$, $\num{0.0704}$] & 0.0 \\
4 & RO-geo & $\num{3.44}$ [$\num{3.10}$, $\num{3.84}$] & $\num{0.0588}$ [$\num{0.0530}$, $\num{0.0659}$] & 0.0 \\
5 & SMC & $\num{337}$ [$\num{310}$, $\num{370}$] & $\num{5.79}$ [$\num{5.32}$, $\num{6.35}$] & 0.0 \\
5 & RO-lin & $\num{4.34}$ [$\num{3.75}$, $\num{5.06}$] & $\num{0.0877}$ [$\num{0.0758}$, $\num{0.102}$] & 0.0 \\
5 & RO-geo & $\num{3.58}$ [$\num{3.03}$, $\num{4.37}$] & $\num{0.0698}$ [$\num{0.0591}$, $\num{0.0853}$] & 0.0 \\
\addlinespace
\multicolumn{5}{l}{$(r_1,r_2)=(1,0.5)$}\\
2 & SMC & $\num{20.0}$ [$\num{18.3}$, $\num{21.9}$] & $\num{0.226}$ [$\num{0.206}$, $\num{0.248}$] & 0.0 \\
2 & RO-lin & $\num{3.73}$ [$\num{3.42}$, $\num{4.10}$] & $\num{0.0517}$ [$\num{0.0473}$, $\num{0.0568}$] & 0.0 \\
2 & RO-geo & $\num{4.02}$ [$\num{3.69}$, $\num{4.42}$] & $\num{0.0527}$ [$\num{0.0483}$, $\num{0.0579}$] & 0.0 \\
3 & SMC & $\num{104}$ [$\num{96.3}$, $\num{115}$] & $\num{1.39}$ [$\num{1.29}$, $\num{1.54}$] & 0.0 \\
3 & RO-lin & $\num{3.71}$ [$\num{3.39}$, $\num{4.06}$] & $\num{0.0603}$ [$\num{0.0551}$, $\num{0.0660}$] & 0.0 \\
3 & RO-geo & $\num{3.87}$ [$\num{3.55}$, $\num{4.25}$] & $\num{0.0603}$ [$\num{0.0553}$, $\num{0.0661}$] & 0.0 \\
4 & SMC & $\num{552}$ [$\num{502}$, $\num{608}$] & $\num{8.35}$ [$\num{7.60}$, $\num{9.20}$] & 0.0 \\
4 & RO-lin & $\num{3.91}$ [$\num{3.27}$, $\num{4.76}$] & $\num{0.0714}$ [$\num{0.0597}$, $\num{0.0868}$] & 0.0 \\
4 & RO-geo & $\num{3.36}$ [$\num{2.78}$, $\num{4.10}$] & $\num{0.0592}$ [$\num{0.0490}$, $\num{0.0724}$] & 0.0 \\
5 & SMC & $\num{2980}$ [$\num{2720}$, $\num{3280}$] & $\num{51.0}$ [$\num{46.5}$, $\num{56.2}$] & 0.0 \\
5 & RO-lin & $\num{6.34}$ [$\num{3.94}$, $\num{12.9}$] & $\num{0.130}$ [$\num{0.0808}$, $\num{0.265}$] & 0.0 \\
5 & RO-geo & $\num{2.33}$ [$\num{0.963}$, $\num{12.0}$] & $\num{0.0478}$ [$\num{0.0197}$, $\num{0.246}$] & 0.0 \\
\addlinespace
\multicolumn{5}{l}{$(r_1,r_2)=(1,0.25)$}\\
2 & SMC & $\num{51.0}$ [$\num{47.0}$, $\num{55.6}$] & $\num{0.571}$ [$\num{0.526}$, $\num{0.622}$] & 0.0 \\
2 & RO-lin & $\num{3.76}$ [$\num{3.45}$, $\num{4.14}$] & $\num{0.0521}$ [$\num{0.0477}$, $\num{0.0573}$] & 0.0 \\
2 & RO-geo & $\num{4.15}$ [$\num{3.82}$, $\num{4.53}$] & $\num{0.0546}$ [$\num{0.0502}$, $\num{0.0596}$] & 0.0 \\
3 & SMC & $\num{300}$ [$\num{274}$, $\num{331}$] & $\num{3.99}$ [$\num{3.64}$, $\num{4.39}$] & 0.0 \\
3 & RO-lin & $\num{3.90}$ [$\num{3.52}$, $\num{4.37}$] & $\num{0.0637}$ [$\num{0.0575}$, $\num{0.0714}$] & 0.0 \\
3 & RO-geo & $\num{3.31}$ [$\num{2.89}$, $\num{3.82}$] & $\num{0.0517}$ [$\num{0.0451}$, $\num{0.0598}$] & 0.0 \\
4 & SMC & $\num{1390}$ [$\num{1270}$, $\num{1530}$] & $\num{21.1}$ [$\num{19.2}$, $\num{23.1}$] & 0.0 \\
4 & RO-lin & $\num{5.45}$ [$\num{3.53}$, $\num{9.67}$] & $\num{0.101}$ [$\num{0.0654}$, $\num{0.180}$] & 0.0 \\
4 & RO-geo & $\num{2.58}$ [$\num{1.50}$, $\num{6.16}$] & $\num{0.0477}$ [$\num{0.0276}$, $\num{0.114}$] & 0.0 \\
5 & SMC & $\num{5180}$ [$\num{4500}$, $\num{5990}$] & $\num{90.2}$ [$\num{78.4}$, $\num{104}$] & 0.0 \\
5 & RO-lin & $\num{1.54}$ [$\num{0.515}$, $\infty$] & $\num{0.0322}$ [$\num{0.0107}$, $\infty$] & 36.3 \\
5 & RO-geo & $\num{18.2}$ [$\num{6.08}$, $\num{8400000000}$] & $\num{0.380}$ [$\num{0.126}$, $\num{176000000}$] & 2.0 \\
\addlinespace
\bottomrule
\end{longtable}}
\subsection{Union experiments}
{\small
\begin{longtable}{@{}clrrr@{}}
\caption{Point estimates and pointwise 95\% percentile bootstrap intervals for union variance gains. $z^*$ is the percentage of zero-variance bootstrap resamples.}\label{tab:boot-union}\\
\toprule
$d$ & Method & $G^{\rm Var}$ [95\% interval] & $G^{\rm CPU}$ [95\% interval] & $z^*$ (\%) \\
\midrule\endfirsthead
\toprule
$d$ & Method & $G^{\rm Var}$ [95\% interval] & $G^{\rm CPU}$ [95\% interval] & $z^*$ (\%) \\
\midrule\endhead
\multicolumn{5}{l}{$(r_1,r_2)=(1,1)$}\\
2 & SMC & $\num{2.69}$ [$\num{2.48}$, $\num{2.95}$] & $\num{0.0380}$ [$\num{0.0346}$, $\num{0.0420}$] & 0.0 \\
2 & RO-lin & $\num{2.14}$ [$\num{1.96}$, $\num{2.36}$] & $\num{0.0357}$ [$\num{0.0326}$, $\num{0.0394}$] & 0.0 \\
2 & RO-geo & $\num{2.42}$ [$\num{2.22}$, $\num{2.63}$] & $\num{0.0388}$ [$\num{0.0356}$, $\num{0.0426}$] & 0.0 \\
3 & SMC & $\num{7.04}$ [$\num{6.45}$, $\num{7.71}$] & $\num{0.0959}$ [$\num{0.0879}$, $\num{0.105}$] & 0.0 \\
3 & RO-lin & $\num{3.02}$ [$\num{2.77}$, $\num{3.30}$] & $\num{0.0487}$ [$\num{0.0446}$, $\num{0.0534}$] & 0.0 \\
3 & RO-geo & $\num{3.68}$ [$\num{3.39}$, $\num{4.01}$] & $\num{0.0575}$ [$\num{0.0530}$, $\num{0.0627}$] & 0.0 \\
4 & SMC & $\num{21.6}$ [$\num{19.9}$, $\num{23.5}$] & $\num{0.323}$ [$\num{0.298}$, $\num{0.353}$] & 0.0 \\
4 & RO-lin & $\num{3.63}$ [$\num{3.33}$, $\num{3.99}$] & $\num{0.0627}$ [$\num{0.0575}$, $\num{0.0688}$] & 0.0 \\
4 & RO-geo & $\num{3.73}$ [$\num{3.42}$, $\num{4.09}$] & $\num{0.0626}$ [$\num{0.0574}$, $\num{0.0687}$] & 0.0 \\
5 & SMC & $\num{63.3}$ [$\num{57.9}$, $\num{69.5}$] & $\num{1.09}$ [$\num{1.00}$, $\num{1.20}$] & 0.0 \\
5 & RO-lin & $\num{3.63}$ [$\num{3.27}$, $\num{4.05}$] & $\num{0.0705}$ [$\num{0.0635}$, $\num{0.0787}$] & 0.0 \\
5 & RO-geo & $\num{3.30}$ [$\num{3.00}$, $\num{3.66}$] & $\num{0.0626}$ [$\num{0.0569}$, $\num{0.0694}$] & 0.0 \\
\addlinespace
\multicolumn{5}{l}{$(r_1,r_2)=(1,0.5)$}\\
2 & SMC & $\num{3.38}$ [$\num{3.11}$, $\num{3.69}$] & $\num{0.0388}$ [$\num{0.0356}$, $\num{0.0423}$] & 0.0 \\
2 & RO-lin & $\num{2.28}$ [$\num{2.09}$, $\num{2.50}$] & $\num{0.0303}$ [$\num{0.0278}$, $\num{0.0333}$] & 0.0 \\
2 & RO-geo & $\num{2.81}$ [$\num{2.58}$, $\num{3.07}$] & $\num{0.0359}$ [$\num{0.0330}$, $\num{0.0394}$] & 0.0 \\
3 & SMC & $\num{9.05}$ [$\num{8.31}$, $\num{9.89}$] & $\num{0.118}$ [$\num{0.107}$, $\num{0.129}$] & 0.0 \\
3 & RO-lin & $\num{3.28}$ [$\num{3.02}$, $\num{3.59}$] & $\num{0.0518}$ [$\num{0.0473}$, $\num{0.0570}$] & 0.0 \\
3 & RO-geo & $\num{3.77}$ [$\num{3.46}$, $\num{4.13}$] & $\num{0.0577}$ [$\num{0.0529}$, $\num{0.0632}$] & 0.0 \\
4 & SMC & $\num{30.5}$ [$\num{28.1}$, $\num{33.4}$] & $\num{0.463}$ [$\num{0.424}$, $\num{0.507}$] & 0.0 \\
4 & RO-lin & $\num{3.56}$ [$\num{3.24}$, $\num{3.92}$] & $\num{0.0638}$ [$\num{0.0580}$, $\num{0.0704}$] & 0.0 \\
4 & RO-geo & $\num{3.57}$ [$\num{3.26}$, $\num{3.91}$] & $\num{0.0618}$ [$\num{0.0564}$, $\num{0.0679}$] & 0.0 \\
5 & SMC & $\num{102}$ [$\num{93.5}$, $\num{111}$] & $\num{1.74}$ [$\num{1.60}$, $\num{1.91}$] & 0.0 \\
5 & RO-lin & $\num{3.86}$ [$\num{3.52}$, $\num{4.27}$] & $\num{0.0753}$ [$\num{0.0686}$, $\num{0.0834}$] & 0.0 \\
5 & RO-geo & $\num{3.16}$ [$\num{2.84}$, $\num{3.53}$] & $\num{0.0601}$ [$\num{0.0540}$, $\num{0.0672}$] & 0.0 \\
\addlinespace
\multicolumn{5}{l}{$(r_1,r_2)=(0.5,0.5)$}\\
2 & SMC & $\num{6.76}$ [$\num{6.24}$, $\num{7.37}$] & $\num{0.0759}$ [$\num{0.0699}$, $\num{0.0827}$] & 0.0 \\
2 & RO-lin & $\num{3.13}$ [$\num{2.88}$, $\num{3.44}$] & $\num{0.0423}$ [$\num{0.0387}$, $\num{0.0465}$] & 0.0 \\
2 & RO-geo & $\num{3.65}$ [$\num{3.34}$, $\num{4.03}$] & $\num{0.0472}$ [$\num{0.0432}$, $\num{0.0521}$] & 0.0 \\
3 & SMC & $\num{38.0}$ [$\num{34.9}$, $\num{41.7}$] & $\num{0.514}$ [$\num{0.473}$, $\num{0.565}$] & 0.0 \\
3 & RO-lin & $\num{3.56}$ [$\num{3.27}$, $\num{3.89}$] & $\num{0.0591}$ [$\num{0.0542}$, $\num{0.0646}$] & 0.0 \\
3 & RO-geo & $\num{4.06}$ [$\num{3.72}$, $\num{4.44}$] & $\num{0.0643}$ [$\num{0.0589}$, $\num{0.0704}$] & 0.0 \\
4 & SMC & $\num{211}$ [$\num{193}$, $\num{232}$] & $\num{3.26}$ [$\num{2.95}$, $\num{3.62}$] & 0.0 \\
4 & RO-lin & $\num{3.77}$ [$\num{3.41}$, $\num{4.19}$] & $\num{0.0684}$ [$\num{0.0613}$, $\num{0.0767}$] & 0.0 \\
4 & RO-geo & $\num{3.17}$ [$\num{2.84}$, $\num{3.57}$] & $\num{0.0553}$ [$\num{0.0491}$, $\num{0.0629}$] & 0.0 \\
5 & SMC & $\num{1190}$ [$\num{1090}$, $\num{1310}$] & $\num{19.8}$ [$\num{18.2}$, $\num{21.7}$] & 0.0 \\
5 & RO-lin & $\num{4.00}$ [$\num{3.17}$, $\num{5.20}$] & $\num{0.0799}$ [$\num{0.0631}$, $\num{0.104}$] & 0.0 \\
5 & RO-geo & $\num{2.66}$ [$\num{1.95}$, $\num{3.82}$] & $\num{0.0520}$ [$\num{0.0381}$, $\num{0.0747}$] & 0.0 \\
\addlinespace
\multicolumn{5}{l}{$(r_1,r_2)=(0.5,0.25)$}\\
2 & SMC & $\num{10.0}$ [$\num{9.21}$, $\num{11.0}$] & $\num{0.115}$ [$\num{0.105}$, $\num{0.126}$] & 0.0 \\
2 & RO-lin & $\num{3.21}$ [$\num{2.96}$, $\num{3.50}$] & $\num{0.0446}$ [$\num{0.0412}$, $\num{0.0486}$] & 0.0 \\
2 & RO-geo & $\num{4.09}$ [$\num{3.74}$, $\num{4.50}$] & $\num{0.0544}$ [$\num{0.0498}$, $\num{0.0599}$] & 0.0 \\
3 & SMC & $\num{57.7}$ [$\num{53.1}$, $\num{63.2}$] & $\num{0.773}$ [$\num{0.711}$, $\num{0.848}$] & 0.0 \\
3 & RO-lin & $\num{3.61}$ [$\num{3.32}$, $\num{3.93}$] & $\num{0.0589}$ [$\num{0.0542}$, $\num{0.0642}$] & 0.0 \\
3 & RO-geo & $\num{4.10}$ [$\num{3.79}$, $\num{4.46}$] & $\num{0.0641}$ [$\num{0.0593}$, $\num{0.0697}$] & 0.0 \\
4 & SMC & $\num{316}$ [$\num{291}$, $\num{345}$] & $\num{4.78}$ [$\num{4.40}$, $\num{5.23}$] & 0.0 \\
4 & RO-lin & $\num{3.70}$ [$\num{3.27}$, $\num{4.25}$] & $\num{0.0675}$ [$\num{0.0596}$, $\num{0.0776}$] & 0.0 \\
4 & RO-geo & $\num{3.20}$ [$\num{2.76}$, $\num{3.71}$] & $\num{0.0559}$ [$\num{0.0482}$, $\num{0.0648}$] & 0.0 \\
5 & SMC & $\num{1680}$ [$\num{1530}$, $\num{1860}$] & $\num{29.8}$ [$\num{27.0}$, $\num{33.0}$] & 0.0 \\
5 & RO-lin & $\num{2.68}$ [$\num{1.96}$, $\num{3.97}$] & $\num{0.0576}$ [$\num{0.0421}$, $\num{0.0854}$] & 0.0 \\
5 & RO-geo & $\num{3.56}$ [$\num{2.44}$, $\num{5.79}$] & $\num{0.0759}$ [$\num{0.0522}$, $\num{0.123}$] & 0.0 \\
\addlinespace
\multicolumn{5}{l}{$(r_1,r_2)=(0.25,0.25)$}\\
2 & SMC & $\num{21.7}$ [$\num{20.0}$, $\num{23.7}$] & $\num{0.247}$ [$\num{0.228}$, $\num{0.269}$] & 0.0 \\
2 & RO-lin & $\num{3.52}$ [$\num{3.24}$, $\num{3.85}$] & $\num{0.0494}$ [$\num{0.0455}$, $\num{0.0541}$] & 0.0 \\
2 & RO-geo & $\num{4.57}$ [$\num{4.20}$, $\num{5.00}$] & $\num{0.0607}$ [$\num{0.0559}$, $\num{0.0664}$] & 0.0 \\
3 & SMC & $\num{163}$ [$\num{150}$, $\num{178}$] & $\num{2.20}$ [$\num{2.03}$, $\num{2.41}$] & 0.0 \\
3 & RO-lin & $\num{3.94}$ [$\num{3.61}$, $\num{4.33}$] & $\num{0.0656}$ [$\num{0.0600}$, $\num{0.0720}$] & 0.0 \\
3 & RO-geo & $\num{3.59}$ [$\num{3.26}$, $\num{3.97}$] & $\num{0.0571}$ [$\num{0.0518}$, $\num{0.0633}$] & 0.0 \\
4 & SMC & $\num{864}$ [$\num{792}$, $\num{949}$] & $\num{13.2}$ [$\num{12.1}$, $\num{14.5}$] & 0.0 \\
4 & RO-lin & $\num{2.99}$ [$\num{2.34}$, $\num{3.89}$] & $\num{0.0564}$ [$\num{0.0441}$, $\num{0.0735}$] & 0.0 \\
4 & RO-geo & $\num{3.53}$ [$\num{2.75}$, $\num{4.77}$] & $\num{0.0649}$ [$\num{0.0506}$, $\num{0.0881}$] & 0.0 \\
5 & SMC & $\num{3580}$ [$\num{3220}$, $\num{4010}$] & $\num{62.5}$ [$\num{56.1}$, $\num{70.1}$] & 0.0 \\
5 & RO-lin & $\num{43.9}$ [$\num{16.5}$, $\infty$] & $\num{0.925}$ [$\num{0.347}$, $\infty$] & 4.9 \\
5 & RO-geo & $\num{3.76}$ [$\num{1.69}$, $\num{26.1}$] & $\num{0.0792}$ [$\num{0.0355}$, $\num{0.550}$] & 0.0 \\
\addlinespace
\multicolumn{5}{l}{$(r_1,r_2)=(0.25,0.125)$}\\
2 & SMC & $\num{27.3}$ [$\num{25.0}$, $\num{29.9}$] & $\num{0.312}$ [$\num{0.286}$, $\num{0.342}$] & 0.0 \\
2 & RO-lin & $\num{3.50}$ [$\num{3.22}$, $\num{3.82}$] & $\num{0.0493}$ [$\num{0.0453}$, $\num{0.0538}$] & 0.0 \\
2 & RO-geo & $\num{4.38}$ [$\num{4.04}$, $\num{4.77}$] & $\num{0.0585}$ [$\num{0.0539}$, $\num{0.0638}$] & 0.0 \\
3 & SMC & $\num{195}$ [$\num{179}$, $\num{214}$] & $\num{2.67}$ [$\num{2.44}$, $\num{2.94}$] & 0.0 \\
3 & RO-lin & $\num{4.04}$ [$\num{3.64}$, $\num{4.49}$] & $\num{0.0680}$ [$\num{0.0613}$, $\num{0.0757}$] & 0.0 \\
3 & RO-geo & $\num{3.55}$ [$\num{3.20}$, $\num{3.96}$] & $\num{0.0571}$ [$\num{0.0515}$, $\num{0.0637}$] & 0.0 \\
4 & SMC & $\num{948}$ [$\num{869}$, $\num{1040}$] & $\num{14.4}$ [$\num{13.2}$, $\num{15.8}$] & 0.0 \\
4 & RO-lin & $\num{4.35}$ [$\num{3.30}$, $\num{6.09}$] & $\num{0.0820}$ [$\num{0.0622}$, $\num{0.115}$] & 0.0 \\
4 & RO-geo & $\num{2.72}$ [$\num{1.91}$, $\num{4.30}$] & $\num{0.0504}$ [$\num{0.0353}$, $\num{0.0799}$] & 0.0 \\
5 & SMC & $\num{3860}$ [$\num{3450}$, $\num{4340}$] & $\num{66.0}$ [$\num{59.1}$, $\num{74.4}$] & 0.0 \\
5 & RO-lin & $\num{136}$ [$\num{45.3}$, $\infty$] & $\num{2.79}$ [$\num{0.926}$, $\infty$] & 37.0 \\
5 & RO-geo & $\num{0.730}$ [$\num{0.247}$, $\num{14300}$] & $\num{0.0150}$ [$\num{0.00506}$, $\num{295}$] & 0.0 \\
\addlinespace
\bottomrule
\end{longtable}}

\end{document}